\documentclass[12pt]{article}

\usepackage{cite}
\usepackage{epsfig}
\usepackage{amsfonts}
\usepackage{amsmath}
\usepackage{amssymb}
\usepackage{tabularx}
\usepackage{axodraw}
\usepackage[dvips]{color}
\usepackage{colordvi,psfrag}
\makeatletter
\def\citer{\@ifnextchar [{\@tempswatrue\@citexr}{\@tempswafalse\@citexr[]}}
 
\def\@citexr[#1]#2{\if@filesw\immediate\write\@auxout{\string\citation{#2}}\fi
  \def\@citea{}\@cite{\@for\@citeb:=#2\do
    {\@citea\def\@citea{--\penalty\@m}\@ifundefined
       {b@\@citeb}{{\bf ?}\@warning
       {Citation `\@citeb' on page \thepage \space undefined}}%
\hbox{\csname b@\@citeb\endcsname}}}{#1}}
\makeatother

\def\refeq#1{\mbox{(\ref{#1})}}
\def\refeqs#1{\mbox{Eqs.~(\ref{#1})}}
\def\reffi#1{\mbox{Fig.~\ref{#1}}}

\def\refta#1{\mbox{Tab.~\ref{#1}}}
\def\refse#1{\mbox{Sect.~\ref{#1}}}

\def\citere#1{\mbox{Ref.~\cite{#1}}}
\def\citeres#1{\mbox{Refs.~\cite{#1}}}

\def\xtilde#1{%
  \setbox0\hbox{$\tilde#1$}%
  \rlap{\raise\ht0\hbox{\tiny$_{\,(\;\,)}$}}%
  \tilde#1%
}

\newcommand{\mste}{m_{\tilde{t}_1}}
\newcommand{\mstz}{m_{\tilde{t}_2}}

\newcommand{\msb}{m_{\tilde{b}}}

\newcommand{\msbe}{m_{\tilde{b}_1}}
\newcommand{\msbz}{m_{\tilde{b}_2}}

\newcommand{\At}{A_t}
\newcommand{\Ab}{A_b}

\newcommand{\msusy}{M_{\text{SUSY}}}

\newcommand{\msqe}{m_{\tilde{q}_1}}
\newcommand{\msqz}{m_{\tilde{q}_2}}

\newcommand{\mhtree}{m_{h,{\rm tree}}}
\newcommand{\mHtree}{m_{H,{\rm tree}}}

\newcommand{\mudim}{\mu^{\drbarm}}

\newcommand{\msbar}{$\overline{\rm{MS}}$}
\newcommand{\msbarm}{\overline{\rm{MS}}}
\newcommand{\drbar}{$\overline{\rm{DR}}$}
\newcommand{\drbarm}{\overline{\rm{DR}}}

\def\order#1{${\cal O}(#1)$}
\newcommand{\cp}{{\cal CP}}
\newcommand{\cD}{{\cal D}}

\newcommand{\cL}{{\cal L}}
\newcommand{\cM}{{\cal M}}
\newcommand{\cU}{{\cal U}}

\newcommand{\edz}{\frac{1}{2}}

\newcommand{\twol}{two-loop}
\newcommand{\onel}{one-loop}

\newcommand{\tc}{{\em TwoCalc}}
\newcommand{\fa}{{\em FeynArts}}
\newcommand{\fh}{{\em FeynHiggs}}

\newcommand{\MW}{M_W}
\newcommand{\MZ}{M_Z}
\newcommand{\MA}{M_A}
\newcommand{\mh}{m_h}
\newcommand{\mH}{m_H}
\newcommand{\Mh}{M_h}
\newcommand{\MH}{M_H}

\newcommand{\mt}{m_{t}}

\newcommand{\mb}{m_{b}}

\newcommand{\mgl}{m_{\tilde{g}}}
\newcommand{\sq}{\tilde{q}}

\newcommand{\sql}{\tilde{q}_L}
\newcommand{\sqr}{\tilde{q}_R}
\newcommand{\sqe}{\tilde{q}_1}
\newcommand{\sqz}{\tilde{q}_2}

\newcommand{\Stop}{\tilde{t}}

\newcommand{\Sbot}{\tilde{b}}
\newcommand{\SbotL}{\tilde{b}_L}

\newcommand{\Sbote}{\tilde{b}_1}
\newcommand{\Sbotz}{\tilde{b}_2}

\newcommand{\tst}{\theta_{\tilde{t}}}
\newcommand{\tsb}{\theta_{\tilde{b}}}

\newcommand{\tsf}{\theta\kern-.20em_{\tilde{f}}}
\newcommand{\tsfp}{\theta\kern-.20em_{\tilde{f}\prime}}
\newcommand{\tsq}{\theta\kern-.15em_{\tilde{q}}}
\newcommand{\sw}{s_\mathrm{w}}
\newcommand{\cw}{c_\mathrm{w}}

\newcommand{\sinQtt}{\sin^2\tst}

\newcommand{\sinQtb}{\sin^2\tsb}
\newcommand{\sinZtb}{\sin 2\tsb}

\newcommand{\sintq}{\sin\tsq}
\newcommand{\sinQtq}{\sin^2\tsq}

\newcommand{\cosQtt}{\cos^2\tst}

\newcommand{\cosQtb}{\cos^2\tsb}
\newcommand{\cosZtb}{\cos 2\tsb}

\newcommand{\costq}{\cos\tsq}
\newcommand{\cosQtq}{\cos^2\tsq}

\newcommand{\KKL}{\left[}
\newcommand{\KKR}{\right]}

\newcommand{\VL}{\left( \begin{array}{c}}
\newcommand{\VR}{\end{array} \right)}
\newcommand{\ML}{\left( \begin{array}{cc}}
\newcommand{\MLd}{\left( \begin{array}{ccc}}
\newcommand{\MLv}{\left( \begin{array}{cccc}}
\newcommand{\MR}{\end{array} \right)}

\newcommand{\re}{\mathop{\rm Re}}
\newcommand{\OP}{\omega_+}
\newcommand{\OM}{\omega_-}

\newcommand{\tb}{\tan \beta}

\newcommand{\sbe}{\sin \beta}
\newcommand{\SQb}{\sin^2\beta\hspace{1mm}}

\newcommand{\Cb}{\cos \beta\hspace{1mm}}
\newcommand{\CQb}{\cos^2\beta\hspace{1mm}}

\newcommand{\Sa}{\sin \alpha\hspace{1mm}}

\newcommand{\Ca}{\cos \alpha\hspace{1mm}}

\newcommand{\gev}{\,\, {\rm GeV}}
\newcommand{\mev}{\,\, {\rm MeV}}

\newcommand{\BC}{\begin{center}}
\newcommand{\EC}{\end{center}}
\newcommand{\BE}{\begin{equation}}
\newcommand{\EE}{\end{equation}}
\newcommand{\BEA}{\begin{eqnarray}}
\newcommand{\BEAnn}{\begin{eqnarray*}}
\newcommand{\EEA}{\end{eqnarray}}
\newcommand{\EEAnn}{\end{eqnarray*}}
\newcommand{\non}{\nonumber}
\newcommand{\id}{{\rm 1\kern-.12em
\rule{0.3pt}{1.5ex}\raisebox{0.0ex}{\rule{0.1em}{0.3pt}}}}
\newcommand{\lsim}
{\;\raisebox{-.3em}{$\stackrel{\displaystyle <}{\sim}$}\;}
\newcommand{\gsim}
{\;\raisebox{-.3em}{$\stackrel{\displaystyle >}{\sim}$}\;}

\newcommand{\gf}{G_F}

\def\al{\alpha}

\def\als{\alpha_s}
\def\alt{\alpha_t}
\def\alb{\alpha_b}

\def\be{\beta}

\def\de{\delta}

\def\De{\Delta}
\def\La{\Lambda}

\def\Si{\Sigma}

\def\hSi{\hat{\Sigma}}

\newcommand{\SLASH}[2]{\makebox[#2ex][l]{$#1$}/}

\newcommand{\pslash}{\SLASH{p}{.2}}

\def\3{\ss}

\begin{document}
\thispagestyle{empty}

\def\thefootnote{\fnsymbol{footnote}}

\begin{flushright}

CERN--PH--TH/2004--126 \hfill
DCPT/04/78 \\
IPPP/04/39 \hfill
MPP--2004--142\\
hep-ph/0411114 \\
\end{flushright}

\vspace{1cm}

\begin{center}

{\Large\sc {\bf High-Precision Predictions}}

\vspace{0.4cm}

{\Large\sc {\bf for the MSSM Higgs Sector at \boldmath{${\cal O}(\alb\als)$}}}

\vspace{1cm}

{\sc 
S.~Heinemeyer$^{1}$%
\footnote{email: Sven.Heinemeyer@cern.ch}%
, W.~Hollik$^{2}$%
\footnote{email: hollik@mppmu.mpg.de}%
, H.~Rzehak$^{2}$%
\footnote{email: hr@mppmu.mpg.de}%
~and G.~Weiglein$^{3}$%
\footnote{email: Georg.Weiglein@durham.ac.uk}
}

\vspace*{1cm}

{\sl
$^1$CERN TH Division, Department of Physics,\\
CH-1211 Geneva 23, Switzerland
 
\vspace*{0.4cm}

$^2$Max-Planck-Institut f\"ur Physik (Werner-Heisenberg-Institut),\\
F\"ohringer Ring 6, D--80805 Munich, Germany

\vspace*{0.4cm}

$^3$Institute for Particle Physics Phenomenology, University of Durham,\\
Durham DH1~3LE, UK

}

\end{center}

\vspace*{1cm}

\begin{abstract}
We evaluate \order{\alb\als} corrections in the Higgs boson
sector of the $\cp$-conserving MSSM, generalising the known result in
the literature to arbitrary 
values of $\tan\beta$. A detailed analysis of the renormalisation in the 
bottom/scalar bottom sector is performed. 
Concerning the lightest MSSM Higgs boson mass, we find relatively
small corrections for positive $\mu$, while 
for $\mu < 0$ the genuine two-loop \order{\alb\als} corrections can
amount up to 3~GeV.
Different renormalisation schemes are applied and numerically compared. 
It is demonstrated that some care has to be taken in choosing an
appropriate renormalisation prescription in order to avoid artificially
large corrections. The residual dependence on the renormalisation scale
is investigated, and the remaining theoretical uncertainties from unknown
higher-order corrections in this sector are discussed for different
regions of the MSSM parameter space.
\end{abstract}

\def\thefootnote{\arabic{footnote}}
\setcounter{page}{0}
\setcounter{footnote}{0}

\newpage


\section{Introduction}

A crucial prediction of the Minimal Supersymmetric Standard Model
(MSSM)~\cite{susy} is the existence of at least one light Higgs
boson. The search for this particle 
is one of the main goals at the present and the next generation
of colliders. Direct searches at LEP have already ruled out a
considerable fraction of the MSSM parameter
space~\cite{LEPHiggsSM,LEPHiggsMSSM}, and the forthcoming
high-energy experiments at the Tevatron, the LHC, and the International
Linear Collider (ILC) will either
discover a light Higgs boson or rule out Supersymmetry (SUSY) 
as a viable theory
for physics at the weak scale. Furthermore, if one or more Higgs
bosons are discovered, bounds on their masses and couplings will be 
set at the LHC~\cite{LHCHiggs,HcoupLHCSMearly,HcoupLHCSM}. 
Eventually the masses and couplings will be determined
with high accuracy at the ILC~\cite{tesla,orangebook,acfarep}. 
Thus, a
precise knowledge of the dependence of masses and mixing angles in
the MSSM Higgs sector on the relevant supersymmetric parameters is of
utmost importance to reliably compare the predictions of the MSSM with
the (present and future) experimental results.

The status of the available results for the higher-order contributions to
the neutral $\cp$-even MSSM Higgs boson masses can be summarised as follows. 
For the
one-loop part, the complete result within the MSSM is 
known~\cite{ERZ,mhiggsf1lA,mhiggsf1lB,mhiggsf1lC}. The 
dominant one-loop contribution is the \order{\alt} term due to top and stop 
loops ($\alt \equiv h_t^2 / (4 \pi)$, $h_t$ being the 
superpotential top coupling).
Corrections from the bottom/sbottom sector can also give large effects,
in particular for 
large values of $\tb$, the ratio of the two vacuum expectation values, 
$\tb = v_2/v_1$. The computation of two-loop corrections is also 
quite advanced. It has now reached a
stage such that all the presumably dominant
contributions are known. They include the strong corrections, usually
indicated as \order{\alt\als}, and Yukawa corrections, \order{\alt^2},
to the dominant one-loop \order{\alt} term, as well as the strong
corrections to the bottom/sbottom one-loop \order{\alb} term 
($\alb \equiv h_b^2 / (4\pi)$), i.e.\ the \order{\alb\als} contribution, 
derived in the limit $\tb \to \infty$.
Presently, the
\order{\alt\als}~\cite{mhiggsEP1b,ahoang,mhiggsRG1,mhiggsRG2,mhiggsletter,mhiggslong,mhiggsEP0,mhiggsEP1,bse,reconc},
\order{\alt^2}~\cite{mhiggsEP1b,ahoang,mhiggsEP3,mhiggsEP2} and the
\order{\alb\als}~\cite{mhiggsEP4} contributions to the self-energies
are known for vanishing external momenta. Most recently also the corrections
\order{\alt\alb} and \order{\alb^2}~\cite{mhiggsEP5}, a ``full'' \twol\
effective potential calculation~\cite{effpotfull} and an evaluation of
the leading two-loop momentum dependent effects~\cite{mhiggsp2} have
become available. In the (s)bottom
corrections the all-order resummation of the $\tb$-enhanced terms,
\order{\alb(\als\tb)^n}, is also performed \cite{deltamb1,deltamb}.
Reviews with further references can be found in 
\citeres{mhiggsAEC,habilSH,Allanach:2004rh}.

\begin{figure}[htb!]
\begin{center}
\epsfig{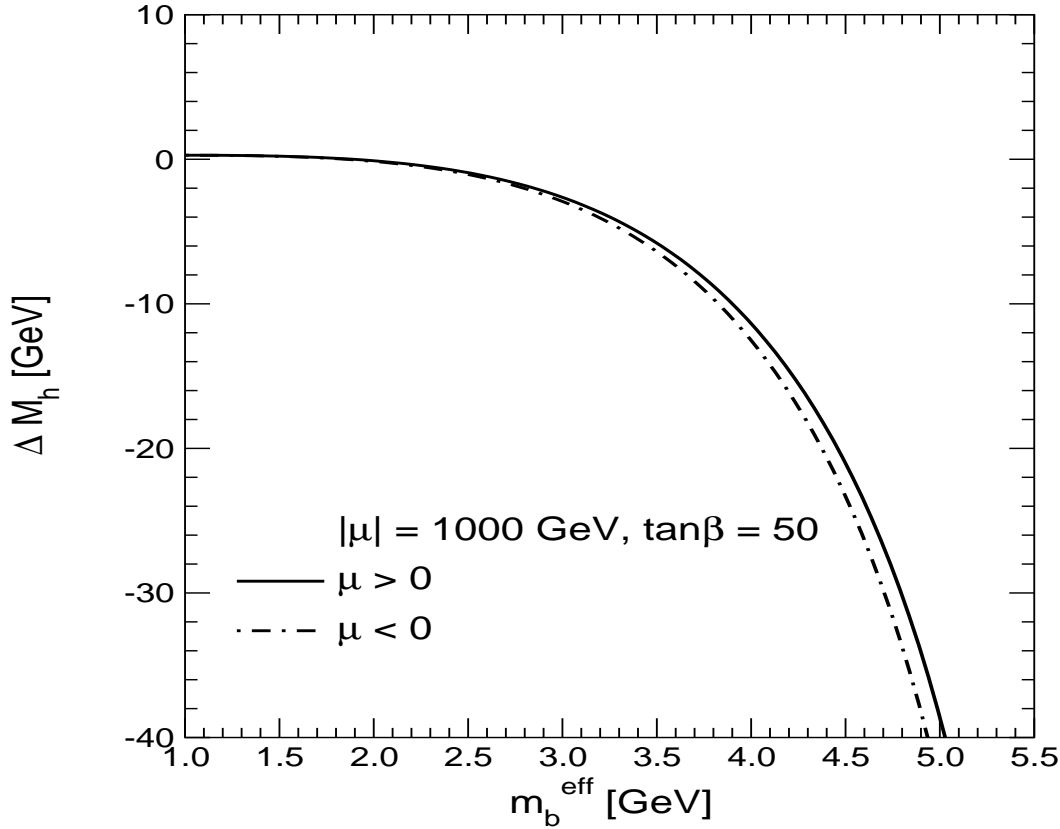}
\caption{
The shift in the lightest $\cp$-even Higgs-boson mass from the 
\onel\ corrections in the $b/\Sbot$~sector is shown as a function of 
the (effective) bottom-quark mass for $\mu = \pm 1000 \gev$, 
$\tan\beta = 50$, $\msusy = 600 \gev$, $\At = \Ab = 500 \gev$, 
$\MA = 700 \gev$.
}
\label{fig:deltamh}
\end{center}
\end{figure}

The $b/\Sbot$ sector has attracted considerable attention in the last years,
since its corrections to the MSSM Higgs boson sector have been found to
be large in certain parts of the MSSM parameter space, possibly even 
exceeding the size of the top/stop corrections.
This can happen especially for large values of $\tb$ and
the supersymmetric Higgs mass parameter $\mu$.
For illustration, we show in \reffi{fig:deltamh}  
the shift in the lightest $\cp$-even Higgs boson mass, $\De\Mh$, arising from
the $b/\Sbot$ sector at the \onel\ level (all \twol\ 
corrections are omitted here) as a function of the bottom-quark mass for 
large $\tb$ and $|\mu|$. The bottom-quark mass in this plot is
understood to be an effective mass that includes higher-order effects
(see the discussion in \refse{sec:ren}). The figure demonstrates that
corrections from the $b/\Sbot$ sector can get large if the effective
bottom mass is bigger than about 3~GeV.

The possibly large size of the corrections from the $b/\Sbot$ sector
makes it desirable to investigate the
corresponding two-loop corrections and thus to analyse
the renormalisation in this sector. An 
inconvenient choice could give rise to artificially large corrections,
whereas a convenient scheme absorbs the dominant contributions into the
one-loop result such that higher-order corrections remain small. The
comparison of different schemes (where no artificially enhanced
corrections appear) gives an indication of the possible size of missing
higher-order terms of \order{\alb\als^2}.

In this paper we derive the result for the \order{\alb\als}
corrections in various renormalisation schemes. The relations between the
different parameters in these schemes are worked out in detail. The
absorption of leading higher-order contributions into an effective
bottom-quark mass is discussed. We perform a numerical
analysis of the various schemes and compare our results
with a previous evaluation of the \order{\alb\als} corrections carried
out in the limit where $\tb$ is infinitely large~\cite{mhiggsEP4}. 
We discuss the dependence of our result on the renormalisation scale and
provide an estimate of the remaining theoretical uncertainties in this
sector.%
\footnote{
This kind of issues have not been addressed in
\citeres{effpotfull,mhiggsp2}.
}

The paper is organised as follows: in \refse{sec:higgssector} we
briefly review the MSSM Higgs boson sector, outline the corresponding
renormalisation at the \twol\ level, and describe the evaluation of the
diagrams of \order{\alb\als}. \refse{sec:ren} contains a detailed
description of the renormalisation of the scalar top and scalar bottom
sector, which is explicitly carried out in four different renormalisation
schemes for the latter. The numerical analysis of the \order{\alb\als}
corrections, the comparison of the different schemes, the investigation
of the renormalisation scale, and the
comparison with the previous result are
performed in \refse{sec:numres}. The conclusions can be found in
\refse{sec:conclusions}.


\section{The Higgs sector at higher orders}
\label{sec:higgssector}

We recall that the Higgs sector of the MSSM~\cite{hhg} comprises two
neutral $\cp$-even Higgs bosons, $h$ and $H$ ($\mh < \mH$), the
$\cp$-odd $A$~boson,%
\footnote{
Throughout this paper we assume that $\cp$ is conserved.
}%
~and two charged Higgs bosons, $H^\pm$.
At the tree-level, the masses $\mhtree$ and $\mHtree$ 
can be calculated in terms of $\MZ$, $\MA$ and $\tb$ from
the mass matrix for the 
neutral $\cp$-even Higgs components (denoted by~$\phi$)
\BEA 
\label{higgsmassmatrixtree}
\cM_\phi &=& \ML \MA^2 \SQb +
\MZ^2 \CQb & -(\MA^2 + \MZ^2) \sbe \Cb \\ -(\MA^2 + \MZ^2) \sbe \Cb &
\MA^2 \CQb + \MZ^2 \SQb \MR   
\EEA
by diagonalization,
\BEA 
\label{higgsmassmatrixtreediag}
\ML \mHtree^2 & 0 \\ 0 & \mhtree^2 \MR &=&  
 \cU_\phi \, \cM_\phi \, \cU_\phi^\dagger~, \qquad
\cU_\phi = \ML \Ca & \Sa \\ -\Sa & \Ca \MR \, ,
\EEA
with the angle $\al$  determined by
\BE
\tan 2\al = \tan 2\beta\, \frac{\MA^2 + \MZ^2}{\MA^2 - \MZ^2},
\quad - \frac{\pi}{2} < \al < 0 . 
\label{alpha}
\end{equation}

\bigskip
In the Feynman-diagrammatic (FD) approach, the higher-order corrected 
Higgs boson masses, $\Mh$ and $\MH$, are derived as the poles of the
$h,H$-propagator matrix,  
i.e.\ by solving the equation
\begin{equation}
\left[p^2 - \mhtree^2 + \hSi_{hh}(p^2) \right]
\left[p^2 - \mHtree^2 + \hSi_{HH}(p^2) \right] -
\left[\hSi_{hH}(p^2)\right]^2 = 0\,.
\label{eq:proppole}
\end{equation}
The renormalised self-energies 
\begin{equation}
\label{SEmatrix}
\hSi(p^2) = \ML \hSi_{HH}(p^2) & \hSi_{hH}(p^2) \\[.2em]
                         \hSi_{hH}(p^2) & \hSi_{hh}(p^2) \MR
\end{equation}
can be expanded according to 
the one-, two-, \ldots loop-order contributions,
\begin{equation}
\label{renSE}
\hSi(p^2) = \hSi^{(1)}(p^2) 
                   + \hSi^{(2)}(p^2) + \cdots ~.
\end{equation}
The dominant \onel\ contributions to the Higgs boson self-energies
(and thus to the Higgs boson masses) from the $b/\Sbot$~sector are
of \order{\alb} and arise from the Yukawa part of the theory (neglecting the
gauge couplings) evaluated at $p^2=0$. This has been verified by comparison
with the full one-loop result from the $b/\Sbot$~sector. 
Hence, the leading \twol\ corrections from the $b/\Sbot$~sector are the
\order{\als} corrections to those dominant \onel\ contributions; they
are obtained in the same limit, i.e.\ for zero external momentum and
neglecting the gauge couplings (the same approximations have been made
in \citere{mhiggsEP4}). This approach is analogous to the way the
leading one- and \twol\ contributions in the top/stop sector have been
obtained, see e.g.~\citere{mhiggslong}.


The renormalisation of the Higgs-boson mass matrix for
the \order{\alb\als} corrections under consideration follows the 
description for the \order{\alt\als} terms given in 
\citere{mhiggslong}.
Renormalisation can be performed by adding the appropriate
counterterms,
\begin{align}\label{massenmatrixrenormierung}
\cM_\phi &\to \cM_\phi + \de\cM^{(1)}_\phi + \de\cM^{(2)}_\phi + \cdots \, ,
\end{align}
 where $\de\cM^{(i)}_\phi$ denotes the {\it i}th-loop 
counterterm matrix consisting of the  
counterterms to the parameters in the tree-level
mass matrix \refeq{higgsmassmatrixtree}.
Field renormalisation is not
needed for the leading \order{\alb\als} corrections.
The renormalised \twol\ Higgs boson self-energies with the leading
contributions of \order{\alb\als} are thus given by
\begin{align}
\label{loopHh}
 \hSi^{(2)}(0) &=
 \Si^{(2)} (0) - \cU_\phi \de \cM_\phi^{(2)} \cU^\dagger_\phi~.
\end{align}

\bigskip
The counterterm matrix in \refeq{loopHh}
is composed of the counterterms for
the $A$-boson mass and for the tadpoles $t_{h,H}$
(with $\sw \equiv \sin\theta_W$, $\cw = \cos\theta_W$),
\begin{align}
\de 
\cM_{\phi}^{(2)} &=
  \begin{pmatrix} \sin^2 \be & - \sin \be \cos
  \be \\ - \sin \be \cos \be & \cos^2 \be \end{pmatrix} \de \MA^{2\, (2)} 
 \nonumber
 \\[1mm]&
\nonumber
\quad\ + \frac{e}{2 \MZ \cw \sw}
 \begin{pmatrix}- \cos \be (1 + \sin^2 \be) &- \sin^3 \be 
              \\ -\sin^3 \be  &\cos \be \sin^2 \be  \end{pmatrix}
(\cos \al \, \de t_H^{(2)} - \sin \al \, \de t_h^{(2)})\\[1mm]&
\quad\  + \frac{e}{2 \MZ \cw \sw}
 \begin{pmatrix}  \cos^2 \be \sin \be&-  \cos^3 \be
 \\-\cos^3 \be & - (1+ \cos^2 \be) \sin \be \end{pmatrix}
 (\sin \al \, \de  t_H^{(2)} + \cos \al \, \de t_h^{(2)}) \, . 
\end{align}

\noindent
The counterterms are determined by the following conditions:
\begin{itemize}
\item[(i)] On-shell renormalisation of the $A$-boson mass, 
  formulated in  the
  approximation of vanishing external momentum, determines the two-loop
  $A$-mass counterterm $\de\MA^{2\,(2)}$ according to
\begin{align}
\de\MA^{2\,(2)} = \Si^{(2)}_{AA} (0) \, .
\end{align}
\item[(ii)] 
  Tadpole renormalisation determines 
  the tadpole counterterms 
  by the requirements
\BEA
\de t_H^{(2)} &= - t_H^{(2)}\, , \quad 
\de t_h^{(2)} &= - t_h^{(2)} \, ,
\EEA
 which means that the minimum of the Higgs potential is not shifted.
\end{itemize}

\begin{figure}[htb!]
\vspace{3em}
\begin{center}
\includegraphics[width=0.9\linewidth]{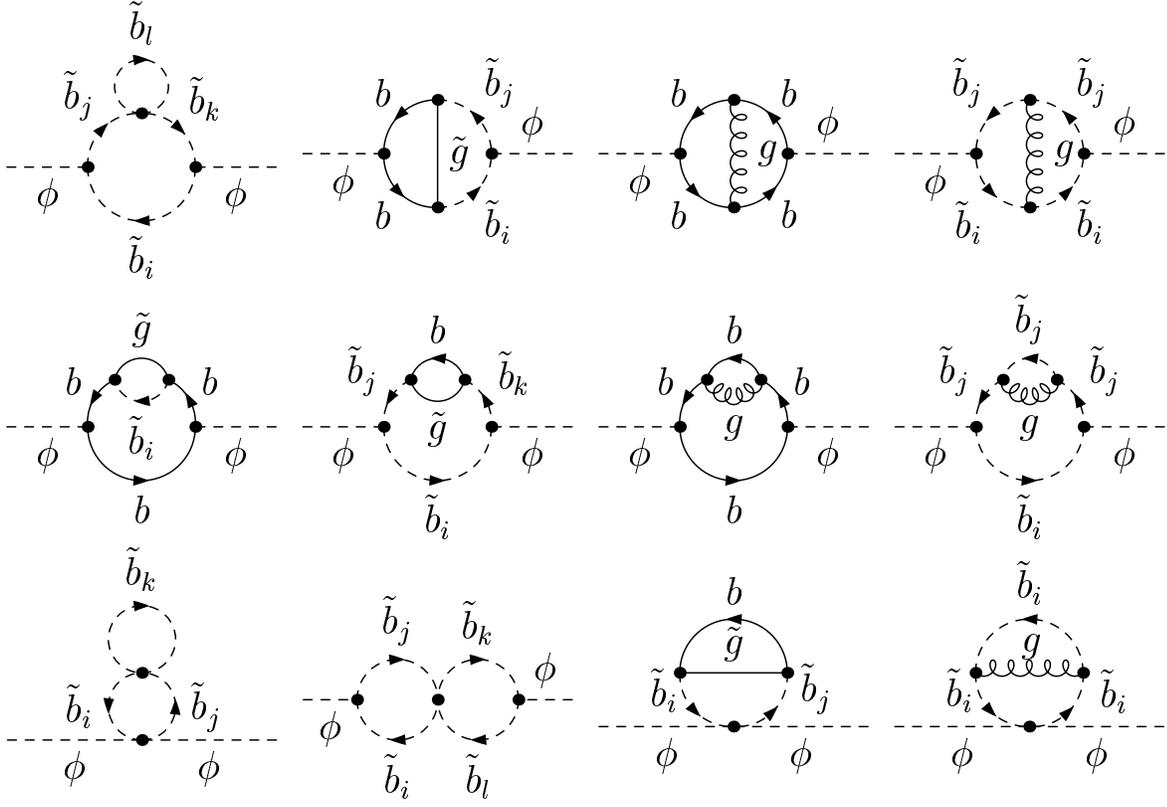}
\vspace{0.5em}
\caption{Generic \twol\ diagrams for the Higgs-boson self-energies
($\phi = h, H, A$;
$\;i,j,k,l = 1,2$).}
\label{fig:FD2L}
\end{center}
\end{figure}

\begin{figure}[htb!]
\vspace{3em}
\begin{center}
\includegraphics[width=0.8\linewidth]{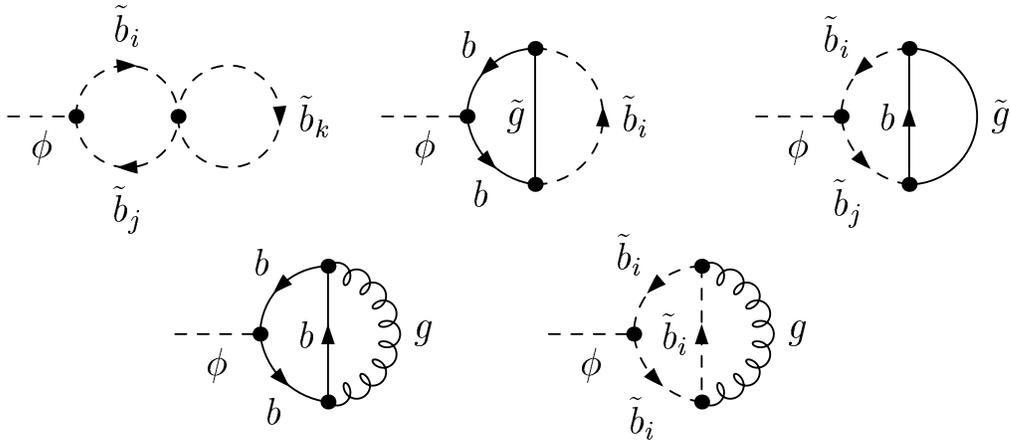}
\caption{Generic \twol\ diagrams for the Higgs tadpoles ($\phi= h,H$;
$\;i,j,k = 1,2$).}
\label{fig:Tad2L}
\end{center}
\end{figure}

\begin{figure}[htb!]
\vspace{3em}
\begin{center}
\includegraphics[width=0.9\linewidth]{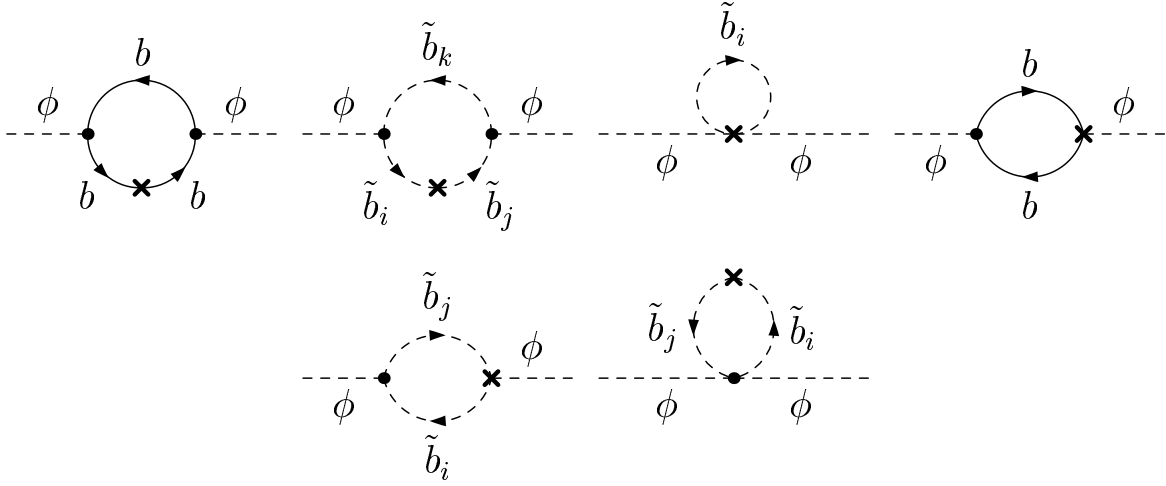}
\vspace{0.5em}
\caption{Generic \onel\ diagrams with counterterm insertion for the
  Higgs-boson self-energies ($\phi = h, H, A$),
$\;i,j,k = 1,2$).}
\label{fig:FD1LCT}
\end{center}
\end{figure}

\begin{figure}[htb!]
\begin{center}
\includegraphics[width=0.9\linewidth]{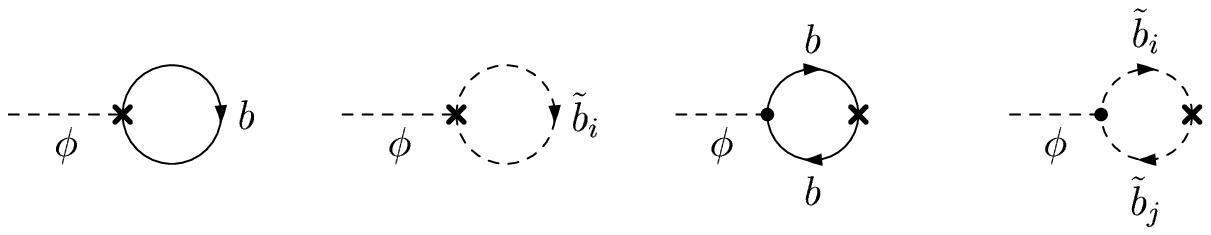}
\caption{Generic one-loop diagrams with counterterm insertion for the
  Higgs tadpoles ($\phi=h,H$, $\;i,j = 1,2$).} 
\label{fig:Tad2Lct}
\end{center}
\end{figure}

The genuine \twol\ Feynman diagrams to be
evaluated for the Higgs boson self-energies and the tadpoles
are shown \reffi{fig:FD2L} and \reffi{fig:Tad2L}.   
The diagrams with subloop renormalisation are depicted in
\reffi{fig:FD1LCT} and \reffi{fig:Tad2Lct}. 
The counterterms for the insertions, where different renormalisation
schemes will be investigated, are specified in the next section.

The diagrams and the corresponding amplitudes have been generated with
the package \fa~\cite{feynarts,famssm}. The further evaluation has
been done using the program \tc~\cite{twocalc}. The resulting expressions
are given in terms of the one-loop functions $A_0$ and
$B_0$~\cite{oneloop}, and the two-loop vacuum integrals~\cite{twoloop}. 


\section{Renormalisation of the quark/squark sector}
\label{sec:ren}

Since the two-loop self-energy is considered at
$\mathcal O(\al_{\{t,\,b\}} \als)$ it is sufficient to
determine the counterterms 
induced by the strong interaction only.

 The squark-mass terms of the Lagrangian, for a given species of
 squarks $\sq$, 
 can be written as the bilinear expression 
\begin{align}
\cL_{\sq\text{-mass}} &= - \begin{pmatrix} \sql^\dagger, \sqr^\dagger 
\end{pmatrix}
\cM_{\sq} \begin{pmatrix} \sql \\ \sqr \end{pmatrix} ,
\end{align} 
with $\cM_{\sq}$ as the squark-mass matrix squared,  
\begin{align}\label{Sfermionmassenmatrix}
\cM_{\sq} = \begin{pmatrix}  
 M_L^2 + m_q^2 + \MZ^2 c_{2 \be} (T_q^3 - Q_q \sw^2) & 
 m_q (A_q - \mu \kappa) \\[.2em]
 m_q (A_q - \mu \kappa) &   
 M_{\sqr}^2 + m_q^2 +\MZ^2 c_{2 \be} Q_q \sw^2 
\end{pmatrix},
\end{align}
where the quantities $M_L^2$, $M_{\sqr}^2$, $A_q$ are 
soft-breaking para\-meters, and $\mu$ is the supersymmetric Higgs mass 
parameter.
Since we are dealing in this paper with a $\cp$-conserving 
Higgs sector, these parameters are treated as real.
As an abbreviation, $c_{2\beta} \equiv \cos(2\beta)$ is introduced;
$\kappa$ is defined as $\kappa = \cot\be$ for {\it up}-type
squarks and $\kappa = \tb$ for {\it down}-type squarks. 
$m_q$, $Q_q$, and $T_q^3$ are mass, charge, and 
isospin of the quark $q$. 
 
The mass matrix \eqref{Sfermionmassenmatrix}
can be diagonalised by a unitary transformation,
which in our case of real parameters 
involves a mixing angle $\tsq$,
\begin{align}
\label{transformation}
\begin{pmatrix} \sqe \\ \sqz \end{pmatrix} =
  \cU_{\sq} \begin{pmatrix} \sql \\ \sqr \end{pmatrix} \, ,
 \qquad 
 \cU_{\sq}  = \begin{pmatrix}U_{\sq_{11}}  & U_{\sq_{12}} \\ 
                           U_{\sq_{21}} & U_{\sq_{22}}  \end{pmatrix} =
 \begin{pmatrix} \cos \tsq & \sin \tsq \\  
                          - \sin \tsq & \cos \tsq \end{pmatrix}\,.
\end{align}
In the $(\sq_1, \sq_2)$-basis, the squared-mass matrix is diagonal,
\begin{align}
\cD_{\sq} &= \cU_{\sq} {\cM}_{\sq}{\cU}_{\sq}^\dagger = 
\begin{pmatrix} \msqe^2 & 0 \\ 0 & \msqz^2 \end{pmatrix}\,,
\end{align}
with the eigenvalues $\msqe^2$ and $\msqz^2$ given by
\begin{align} \non
&m_{\sq_{1,2}}^2 = \edz (M_L^2 +M_{\sqr}^2) +  m_q^2 + 
                         \edz T_q^3 \MZ^2 c_{2 \be} 
\pm \edz \frac{M_L^2 - M_{\sqr}^2+ \MZ^2 c_{2 \be} (T_q^3 - 2 Q_q
\sw^2)}{|M_L^2 - M_{\sqr}^2+ \MZ^2 c_{2 \be} (T_q^3 - 2 Q_q\sw^2)|}
 \\& \qquad\ \quad 
                  \times      \sqrt{\bigl[ M_L^2 - M_{\sqr}^2  
                      + \MZ^2 c_{2 \be} (T_q^3 - 2 Q_q \sw^2)\bigr]^2 
                      + 4  m_q^2 (A_q - {\mu} \kappa)^2}   \; .
\label{Sfermionmassenmatrixeigenwerte}\end{align}
The squark-mass matrix  can now be expressed in terms of the two mass
eigenvalues and the mixing angle, yielding
\begin{equation}
\label{stopmassenmatrix}
\cM_{\sq} = \ML \cosQtq \; \msqe^2 + \sinQtq \; \msqz^2 &
                \sintq \costq \, (\msqe^2 - \msqz^2) \\[.4em]
                \sintq \costq \, (\msqe^2 - \msqz^2) &
                \sinQtq \;\msqe^2 + \cosQtq \;\msqz^2 \MR~.
\end{equation}


\subsection{Renormalisation of the top and scalar top sector }
\label{subsec:stoprenorm}

The $(t,\tilde{t})$ sector contains four independent parameters:
the top-quark mass $\mt$, the stop masses $\mste$ and $\mstz$, and
either the squark mixing angle $\tst$ or, equivalently, the
trilinear coupling $\At$. Accordingly, the renormalisation of this sector 
is performed by introducing four counterterms that are determined by
four independent renormalisation conditions.

The following renormalisation conditions are imposed
(the procedure is equivalent to that of \citere{hr}, although
there no reference is made to the mixing angle).  
\begin{itemize}
\item[(i)] 
On-shell renormalisation of the top-quark mass yields the top
mass counterterm,
\begin{align}\label{dmt}
\de\mt = \edz \mt \bigl [\re \Si_{{t}_L} (\mt^2) +
                         \re \Si_{{t}_R} (\mt^2) + 
                        2\re \Si_{{t}_S} (\mt^2)
\bigr]\; ,
\end{align}
with the scalar coefficients of the
unrenormalised top-quark self-energy, $\Si_t (p)$, 
in the Lorentz decomposition
\begin{align}
\label{Fermionselbstenergiezerlegung}
 \Si_t (p) &= \pslash \OM
\Si_{{t}_L} (p^2) + \pslash \OP
\Si_{{t}_R} (p^2) + \mt \Si_{{t}_S} (p^2)\; .
\end{align}
\item[(ii)] 
On-shell renormalisation of the stop masses
determines the mass counterterms
\begin{equation}
\label{dmst}
\de \mste^2 = 
\re \Si_{\Stop_{11}}(m_{{\Stop}_{1}}^2) \, , \quad
\de \mstz^2 = 
\re \Si_{\Stop_{22}}(m_{{\Stop}_{2}}^2)\; ,
\end{equation}
in terms of the diagonal squark self-energies.
\item[(iii)] 
The counterterm for the mixing angle, $\tst$, (entering
\refeq{stopmassenmatrix}) is fixed in the following way,
\begin{align}
\label{ZusammenhangdeltathetadeltaM}  
\de \tst = 
\frac{\re \Si_{\Stop_{12}}(m_{{\Stop}_{1}}^2)+\re 
\Si_{\Stop_{12}}(m_{{\Stop}_{2}}^2)}{2(\mste^2-\mstz^2)}\; ,
\end{align} 
involving the non-diagonal squark self-energy. (This is a convenient
choice for the treatment of \order{\als} corrections. If
electroweak contributions were included, a manifestly gauge-independent
definition would be more appropriate.)
\end{itemize}
In the renormalised vertices with squark and Higgs fields,
the counterterm of the trilinear
coupling $\At$ appears. 
Having already specified $\de \tst$,
the $A_t$ counterterm cannot be defined
independently but follows from the relation
\begin{align}
\sin 2 \tst = \frac{ 2 \mt ( \At - \mu \cot \be)}{\mste^2 -\mstz^2} \, ,
\end{align} 
yielding 
\begin{align}
\de \At &=  \frac{1}{\mt}
 \Bigl[\frac{1}{2} \sin 2 \tst
 \bigl(\de \mste^2 - \de \mstz^2\bigr) 
      + \cos 2 \tst
      (\mste^2 - \mstz^2)\, \de \tst 
\non \\[1.5mm] & \quad\ \quad\ \label{deltaAt}
   - \frac{1}{2 \mt} \sin 2 \tst  (\mste^2 -\mstz^2) \, \de \mt \Bigr]~.
\end{align}
This relation is valid at \order{\als} since both $\mu$ and $\tan\beta$
do not receive one-loop contributions from the strong interaction.


\subsection{Renormalisation of the bottom and scalar bottom sector}
\label{subsec:sbotrenorm}

Because of SU(2)-invariance 
the soft-breaking parameters for the left-handed {\em up}- and
{\em down}-type squarks are identical, and thus
the squark masses of a given generation are not
independent. The stop and sbottom masses are connected via the relation
\begin{align}
 \cosQtb \msbe^2 + \sinQtb \msbz^2 = 
 \cosQtt \mste^2 + \sinQtt \mstz^2 +
\mb^2 - \mt^2 - M_W^2 \cos (2 \beta)~,
\label{MSb1gen}
\end{align}
with the entries of the rotation matrix in \refeq{transformation}.  
Since the stop masses have already been renormalised
on-shell, only one of the sbottom mass
counterterms can be determined independently. In the following, the
$\Sbotz$~mass is chosen%
\footnote{
 This choice is possible since 
 \eqref{transformation}--\eqref{Sfermionmassenmatrixeigenwerte} 
 ensure that
 the $\Sbotz$-field and the $\SbotL$-field do not coincide.
}%
~as the pole mass yielding the counterterm from an on-shell
renormalisation condition, i.e.\ 
\begin{align}
\de \msbz^2 &= \re \Si_{\Sbot_{22}}(\msbz^2)\; ,
\label{eq:msbz}
\end{align}
whereas the counterterm for $\msbe$ is determined 
as a combination of other counterterms, 
according to
\begin{align} 
\de \msbe^2 &= 
\frac{1}{\cos^2 \tsb} \Bigl(
  \cos^2 \tst \de \mste^2   
 + \sin^2 \tst \de \mstz^2 
 - \sin^2 \tsb \de \msbz^2
 - \sin 2 \tst (\mste^2 -\mstz^2)\de \tst \nonumber
 \\[1.5mm]
& \quad 
 + \sin 2 \tsb (\msbe^2 -\msbz^2)\de \tsb 
-  2 \mt\, \de\mt + 2 \mb\, \de \mb   \Bigr)~.
\label{ms1CT}
\end{align}
Accordingly, the numerical value of $\msbe$ does not
correspond to the pole mass. The pole mass can be obtained 
from $\msbe$ via a finite shift of $\mathcal O(\als)$
(see e.g.~\citere{delrhosusy2loop}).

There are three more parameters with counterterms to be determined: 
the $b$-quark mass $\mb$, the mixing angle $\tsb$,
and the trilinear coupling~$\Ab$. They are connected via
\begin{align}
\label{mixingangleAparametermbrelation}
\sin 2 \tsb = \frac{ 2 \mb ( \Ab - \mu\tb)}{\msbe^2 -\msbz^2} \, ,
\end{align} 
which reads in terms of counterterms
\begin{align}
\label{deltaSbot}
2 \cosZtb\; \de\tsb &= \sinZtb \frac{\de\mb}{\mb} 
                   + \frac{2\mb\,\de\Ab}{\msbe^2 - \msbz^2}
                   - \sinZtb \frac{\de\msbe^2 - \de\msbz^2}
                                  {\msbe^2 - \msbz^2}~.
\end{align}
Only two of the three counterterms, $\de\mb$, $\de\tsb$, $\de\Ab$ can
be treated as independent, which offers a variety of choices.
In the following,
four different renormalisation schemes, see \refta{tab:sbotren},
will be investigated.
Two of them are on-shell schemes
in the sense that the Higgs self-energies do not depend on the
renormalisation scale $\mu^{\drbarm}$. 

\begin{table}[!htb]
\renewcommand{\arraystretch}{1.7}
\begin{tabularx}{16.5cm}{|c||X|X|X|X|}
\hline 
scheme & $\msbz^2$ & $\mb$  & $\Ab$ & $\tsb$ \\ \hline\hline
analogous to $t/\Stop$ sector (``$\mb$ OS'') & on-shell & 
on-shell  & & on-shell\\\hline
\drbar\ bottom-quark mass  (``$\mb$ \drbar'') & on-shell &
\drbar   & \drbar  & \\\hline 
\drbar\ mixing angle and $\Ab$ (``$\Ab$, $\tsb$ \drbar'') & on-shell &   & 
\drbar  & \drbar  \\\hline 
on-shell mixing angle and $\Ab$ (``$\Ab$, $\tsb$ OS'') & on-shell &   &
on-shell & on-shell \\\hline 
\end{tabularx}
\caption {\small Summary of the four renormalisation schemes for the bottom
quark/squark sector investigated below. Blank entries indicate
dependent quantities.} 
\label{tab:sbotren}
\end{table} 

The schemes are described in the following subsections, prior to the
discussion of their quantitative numerical features in~\refse{sec:numres}.


\subsubsection{Analogous to the top quark/squark sector}

A straight-forward 
possibility is to impose renormalisation conditions in analogy to those
of the top quark/squark sector in \refse{subsec:stoprenorm}.
\begin{itemize}
\item[(i)] 
On-shell renormalisation of the bottom quark mass $\mb$
  determines the corresponding counterterm as follows,
\begin{align}
\de \mb = \edz  \mb \bigl[\re {\Si}_{{b}_L} (\mb^2) +
\re  {\Si}_{{b}_R} (\mb^2) + 2\re  {\Si}_{{b}_S} (\mb^2) \bigr] \, .
\label{eq:dembos}
\end{align}
\item[(ii)] 
The counterterm for the sbottom mixing angle
$\tsb$ is determined in the following way,
\begin{align} \label{dthetab}
\de \tsb = 
\frac{\re \Si_{\tilde{b}_{12}}(\msbe^2)
     +\re \Si_{\tilde{b}_{12}}(\msbz^2)}
     {2(\msbe^2 - \msbz^2)}\; .
\end{align} 
\end{itemize}
The dependent counterterm $\de \msbe^2$ for the $\tilde{b}_1$ mass
is then fully specified by~\refeq{ms1CT}. Moreover, 
$\Ab$ is treated here as a dependent quantity; 
the corresponding counterterm
$\de\Ab$ follows from the relation \refeq{deltaSbot}, 
yielding in combination with (\ref{ms1CT}) the expression
\begin{align} \non
\de \Ab &= \frac{1}{\mb}\Bigl[
-\tan \tsb \de \msbz^2 + (\msbe^2-\msbz^2) \de \tsb
- \de m_{b} \Bigl( \frac{1}{2 \mb}(\msbe^2-\msbz^2) \sin 2 \tsb
             - 2 \tan \tsb \mb \Bigr) \\[1.5mm] 
\label{Abparameter}
& \quad\ \quad\
+ \tan \tsb \Bigl(\cos^2 \tst \de \mste^2
 +  \sin^2 \tst \de \mstz^2 
 -   \sin 2 \tst
         (\mste^2-\mstz^2) \de \tst 
-2 \mt \de \mt\Bigr) \Bigr]~.
\end{align}

While formally the renormalisation described in this section is the same as 
in the top/stop sector, there are nevertheless important differences. 
The top-quark pole mass can be directly extracted from experiment and,
due to its large numerical value as compared to other quark masses 
and the fact that
the present experimental error is much larger than the QCD scale, 
it can be used as input for theory predictions in a well-defined way.
For the mass of the bottom quark, on the other hand, problems related to
non-perturbative effects are much more severe.
Therefore the parameter extracted from the comparison of theory and 
experiment~\cite{pdg} 
is not the bottom pole mass. Usually the value of the bottom mass 
is given in the \msbar\ renormalisation scheme,
with the renormalisation scale $\mu^{\msbarm}$ chosen as the
bottom-quark mass, i.e.\ $\mb^{\msbarm}(\mb^{\msbarm})$~\cite{pdg}.

Another important difference to the top/stop sector is the replacement
of $\cot\be \rightarrow \tan\be$. As will be discussed in more detail below, 
very large effects can occur in this scheme for large values of $\mu$
and $\tan\be$.


\subsubsection{\boldmath{\drbar} bottom-quark mass}

Potential problems with the bottom pole mass can be avoided by adopting
a renormalisation scheme with a running bottom-quark mass. In the
context of the MSSM it seems appropriate to use the \drbar\
scheme~\cite{dred} and to include the SUSY contributions at 
$\mathcal O(\als)$ into the running. We therefore choose a scheme where 
$\mb$ and $A_b$ are both renormalised in the \drbar\ scheme.
The following renormalisation conditions are imposed 
for the independent quantities.
\begin{itemize}
\item[(i)] 
The $b$-quark mass is defined in the \drbar\ scheme, which determines
the mass counterterm by the expression
 \begin{align}
 \de \mb = \edz  \mb \bigl[
 \re {\Si}_{{b}_L}^{\rm div} (\mb^2)  +
 \re  {\Si}_{{b}_R}^{\rm div} (\mb^2) + 
 2\re {\Si}_{{b}_S}^{\rm div} (\mb^2) 
 \bigr]\; ,
\label{eq:dembdrbar}
 \end{align}
where ${\Si}^{\rm div}$ means replacing the one- and two-point integrals
$A$ and $B_0$ in the quark self-energies
by their divergent parts in the following way,
 \begin{align}
\nonumber
 A(m)|_{\rm div} &= m^2 \De \; , \\ 
\label{Ersetzung}
 B_0(p^2, m_1, m_2)|_{\rm div} &= \De\;, 
 \end{align} 
with $\De = 2/\epsilon-\gamma+\log 4\pi$, and $D = 4 - \epsilon$.
\item[(ii)] 
Besides $\mb$, also the trilinear coupling $\Ab$ is 
defined within the \drbar\ scheme. Using \refeq{Abparameter} and
inserting the self-energies yields the counterterm
 \begin{align} \non
 \de \Ab &= 
 \frac{1}{\mb}\Bigl[
 -\tan \tsb
   \re \Si_{\Sbot_{22}}^{\rm div}(\msbz^2)
 + \frac{1}{2}
  (\re \Si_{\Sbot_{12}}^{\rm div}(\msbe^2)+
   \re \Si_{\Sbot_{12}}^{\rm div}(\msbz^2))\\[1.5mm]
& \non \quad\ \quad\ 
  +  \tan \tsb \Bigl(\cos^2 \tst
  \re  \Si_{\Stop_{11}}^{\rm div}(\mste^2)
  + \sin^2 \tst
  \re  \Si_{\Stop_{22}}^{\rm div}(\mstz^2) \\[1.5mm]
& \non \quad\ \quad\ 
 - \frac{1}{2}  \sin 2 \tst
  (\re \Si_{\Stop_{12}}^{\rm div}(\mste^2)+
 \re \Si_{\Stop_{12}}^{\rm div}(\mstz^2)) \Bigr) 
\\[1.5mm]
& \non \quad\ \quad\
 -  \mt^2 
 \Bigl(\re {\Si}_{{t}_L}^{\rm div}(\mt^2) + 
  \re {\Si}_{{t}_R}^{\rm div}(\mt^2) + 
 2\re {\Si}_{{t}_S}^{\rm div}(\mt^2)\Bigr)
\Bigr] \\[1.5mm]
& \non \quad\ 
  + \edz  \bigl(2 \tan \tsb \mb - \frac{1}{2 \mb}(\msbe^2-\msbz^2) \sin 2
 \tsb \bigr)\\[1.5mm] 
& \quad \quad
  \Bigl 
  (\re {\Si}_{{b}_L}^{\rm div} (\mb^2)
 + \re {\Si}_{{b}_R}^{\rm div} (\mb^2) 
 + 2\re {\Si}_{{b}_S}^{\rm div} (\mb^2) \Bigr)~.   
\label{AbcountertermDR}
 \end{align} 
\end{itemize}
The counterterms for the mixing angle, $\de \tsb$, and
the $\tilde{b}_1$ mass, $\de \msbe^2$, 
are dependent quantities and can be determined
as combinations of the independent counterterms,
invoking (\ref{ms1CT}) 
and~(\ref{deltaSbot}),
\begin{align}\non
 \de \tsb &= \frac{1}{\msbe^2 -\msbz^2} \Bigl[
  \mb \de \Ab 
 +\tan \tsb \de \msbz^2
 + \de\mb
  \Bigl(\frac{1}{2 \mb}(\msbe^2 -\msbz^2) \sin 2 \tsb  - 2 \tan \tsb
  \mb\Bigl)  
\\[1.5mm] &   \quad\  \label{thetabinAbMBdrbar}   
- \tan \tsb \Bigl(\cos^2 \tst \de\mste^2
 + \sin^2 \tst \de\mstz^2 
 -    \sin 2 \tst (\mste^2 -\mstz^2) \de \tst 
 - 2 \mt \de \mt \Bigr) \Bigr] ~,
\\[2mm]
%
\de \msbe^2 &= \tan^2 \tsb \de \msbz^2 
+ 2 \tan \tsb \mb\de\Ab 
+ 2 \Bigl( \frac{1}{\mb} \sin^2 \tsb (\msbe^2-\msbz^2)
        + (1 - \tan^2 \tsb) \mb \Bigr) \de \mb \nonumber
\\[1mm]
 &   \quad \label{msb1inAbMBdrbar}
+(1 - \tan^2 \tsb)  
 \Bigl(\cos^2 \tst \de \mste^2 
      + \sin^2 \tst \de \mstz^2 
      - \sin 2 \tst (\mste^2-\mstz^2) \de \tst - 2 \mt \de \mt  \Bigr)~. 
\end{align}

The renormalised quantities in this scheme depend on the \drbar\
renormalisation scale $\mu^{\drbarm}$.
If not stated differently, in all numerical results given in this paper the
\drbar\ scale refers to the top-quark mass, i.e.\ 
$\mu^{\drbarm} = m_t$.

\bigskip
In order to determine the value of 
$\mb^{\drbarm, \text{MSSM}}(\mu^{\drbarm})$ from the 
value $\mb^{\msbarm}(\mu^{\msbarm})$ that is extracted from the experimental
data one has to note that by definition $\mb^{\drbarm, \text{MSSM}}$ contains 
all MSSM contributions at \order{\als}, while $\mb^{\msbarm}$ contains
only the \order{\als} SM correction, i.e.\ the gluon-exchange
contribution. Furthermore, a finite shift arises from the transition
between the \msbar\ and the \drbar\ scheme. As input value for 
$\mb^{\msbarm}(M_Z)$ we use in this paper $\mb^{\msbarm}(M_Z) =
2.94$~GeV\cite{mbmsbarmz}.

The expression for $\mb^{\drbarm, \text{MSSM}}(\mu^{\drbarm})$ 
is most easily derived
by formally relating $\mb^{\drbarm, \text{MSSM}}$ to the bottom pole mass 
first and
then expressing the bottom pole mass in terms of the \msbar\ mass (the
large non-perturbative contributions affecting the bottom pole mass drop 
out in the relation of $\mb^{\drbarm, \text{MSSM}}$ to $\mb^{\msbarm}$).
Using the equality $\mb^{\rm OS} + \de \mb^{\rm OS} = \mb^{\drbarm,
\text{MSSM}} + \de \mb^{\drbarm, \text{MSSM}}$ and the expressions for 
the on-shell counterterm and
the \drbar\ counterterm given in \refeq{eq:dembos} and \refeq{eq:dembdrbar}, 
respectively, one finds
\begin{equation}
\label{eq:mbdrbar1}
\mb^{\drbarm, \text{MSSM}}(\mu^{\drbarm}) = \mb^{\text{OS}} + \edz  \mb
  \bigl(\Si^{\rm fin}_{b_L}({\mb}^2) + \Si^{\rm fin}_{{b}_R} ({\mb}^2) \bigr) 
+ \mb\, \Si^{\rm fin}_{b_S}(\mb^2)~.
\end{equation}
Here the $\Si^{\rm fin}$ are the UV-finite parts of the self-energy
coefficients in \refeq{eq:dembos}. 
They depend on the \drbar\ scale 
$\mu^{\drbarm}$ and are evaluated for
on-shell momenta, $p^2 = \mb^2$. 
Inserting $\mb^{\text{OS}} =
\mb^{\msbarm}(M_Z) b^{\text{shift}}$, where
\begin{align}\label{dregdredbmassshift}
b^{\text{shift}} \equiv \Bigl[1 +
\frac{\alpha_s}{\pi} \Bigl(\frac{4}{3} - \ln
\frac{(\mb^{\msbarm})^2}{M_Z^2} \Bigr)\Bigr]~,   
\end{align}
one finds the desired expression for $\mb^{\drbarm}$,
\begin{align}
  \label{eq:mbdrbar2}
\mb^{\drbarm, \text{MSSM}}(\mu^{\drbarm}) &= \mb^{\msbarm}(M_Z)
  b^{\text{shift}} + \edz  \mb 
  \Bigl(\Si^{\rm fin}_{b_L}({\mb}^2) + \Si^{\rm fin}_{{b}_R} ({\mb}^2)
\Bigr)
+ \mb\, \Si^{\rm fin}_{b_S}(\mb^2)~.
\end{align}


\subsubsection{\boldmath{\drbar} mixing angle and \boldmath{$\Ab$}} 

A further possibility is to impose renormalisation conditions for the
mixing angle $\tsb$ and for $\Ab$, and to treat the
counterterm of the $b$-quark mass as a dependent quantity determined 
as a combination of the other counterterms using the relation
\refeq{deltaSbot}.
The renormalisation conditions in this case read explicitly:
\begin{itemize}
\item[(i)] 
$\de\Ab$ is determined in the \drbar\ scheme as in the previous case
by the expression~\refeq{AbcountertermDR}.
\item[(ii)] 
The mixing angle $\tsb$, defined in the \drbar\ scheme,
 is renormalised by the counterterm
 \begin{align}\label{thetadrbar} 
 \de \tsb = 
 \frac{\re \Si_{\Sbot_{12}}^{\rm div}(\msbe^2)+
       \re \Si_{\Sbot_{12}}^{\rm div}(\msbz^2)}
      {2(\msbe^2-\msbz^2)} \; .
 \end{align}
\end{itemize}
The counterterm for the $b$-quark mass, $\de \mb$, can be obtained
using \refeq{deltaSbot}
and the constraint~\refeq{ms1CT}.
It is given by the following
quantity (which is well-behaved for $\tsb \to 0$),
\begin{align}
\de \mb &= \Bigl[\tan \tsb  \Bigl(- \de \msbz^2
+ \cos^2 \tst \de \mste^2 + \sin^2 \tst \de \mstz^2   - \sin 2 \tst
  (\mste^2 -\mstz^2)\de \tst - 2 \mt\, \de \mt\Bigr)
 \non \\[1.5mm]\label{deltambabh}
& \quad\ - \mb\, \de \Ab + 
 (\msbe^2 - \msbz^2)  \de \tsb
  \Bigr]
 \Bigl[\frac{\msbe^2 -\msbz^2}{ 2 \mb} \sin 2 \tsb 
 - 2 \mb \tan \tsb \Bigr]^{-1} ~.
\end{align}
The numerical value of $\mb$ in this scheme is obtained from 
\refeq{eq:mbdrbar2} and the (finite) difference of the counterterms
given in \refeq{deltambabh} and \refeq{eq:dembdrbar}.

Finally, \eqref{ms1CT} yields also 
the counterterm for the dependent squark mass, $\de \msbe^2$,
with the specification~\refeq{deltambabh}
for the $b$-mass counterterm. 


\subsubsection{On-shell mixing angle and \boldmath{$\Ab$} }
\label{subsubsec:AbtsbOS}

In \citere{mhiggsEP4} a renormalisation condition was imposed on the 
$A\Sbote\Sbotz$ vertex in order to avoid an explicit dependence on the
renormalisation scale $\mu^{\drbarm}$. For the purpose of comparing our 
results with those of \citere{mhiggsEP4} we include such a renormalisation
scheme in our discussion. While in \citere{mhiggsEP4} 
the limit $\tb \to \infty$ has been used to
derive all the renormalisation conditions and counterterms, we have
derived the relevant quantities for arbitrary values of $\tb$. 
We call this scheme ``on-shell'' (as in \citere{mhiggsEP4}), although 
the vertex is taken at an off-shell value of the $A$-boson momentum.

Similarly to the previous scheme,
the counterterm for the $b$-quark mass is derived as
a linear combination of other counterterms by means 
of~\refeq{deltaSbot}.
The independent renormalisation conditions can be formulated as follows.
\begin{itemize}
\item[(i)] 
The counterterm for the mixing angle $\tsb$ is defined by 
\begin{align}\label{mixangleUIF} 
\de \tsb = 
  \frac{\re \Si_{\Sbot_{12}}(\msbe^2)+
        \re \Si_{\Sbot_{12}}(\msbz^2)}{2(\msbe^2-\msbz^2)}\; ,
\end{align} 
as in the scheme ``analogous to the top quark/squark sector''.
\item[(ii)] 
$\Ab$ is determinded by imposing the condition
\begin{align}\label{lambdahutbed}
\hat{\La}(0, \msbe^2, \msbe^2) + \hat{\La}(0, \msbz^2, \msbz^2) = 0\; ,
\end{align}
with $\hat{\La}(p_A^2, p_{\Sbote}^2, p_{\Sbotz}^2)$ as the
renormalised three-point $A\Sbote\Sbotz$
vertex function, \\
\BC
\vspace{-3em}
\setlength{\unitlength}{1pt}
\begin{picture}(260, 180)
\DashArrowLine(140,090)(220,130){5}
\DashArrowLine(220,050)(140,090){5}
\DashLine(50,90)(140,90){5}
\put(35,85){$A$}
\put(225,45){$\Sbotz$}
\put(225,130){$\Sbote$}
\put(240,85){$\Hat{=} \; \hat{\La}(p_A^2, p_{\Sbote}^2, p_{\Sbotz}^2)$~,}\,
\GCirc(140,90){10}{.5}
\end{picture}
\vspace{-3em}
\EC
\begin{align}\non
 \hat{\La}(p_A^2, p_{\Sbote}^2, p_{\Sbotz}^2) &= \La (p_A^2,
 p_{\Sbote}^2, p_{\Sbotz}^2) + \frac{i e}{2 M_W \sin \theta_W}\Bigl[
 \mb \tb\, \de\Ab  \\ & \quad  + (\mu + \tb \Ab)\Bigl( \de \mb +
 \frac{1}{2} \mb (\de
 Z_{\Sbote  \Sbote} + \de Z_{\Sbotz \Sbotz})\Bigr)\Bigr] .
\label{lambdahut}
\end{align}
In the large-$\tb$ limit, 
this requirement reproduces the
condition applied in \citere{mhiggsEP4}. 

\noindent
Condition (ii) can be formulated as an equation determining the counterterm for
$A_b$ in the  following way,
\begin{align}
\non
\de \Ab &= 
  i \frac{\MW \sin \theta_W}{e\, \mb \tb} 
 \Bigl( \La(0, \msbe^2, \msbe^2) + \La(0, \msbz^2, \msbz^2) \Bigr)
- \frac{\mu + \Ab \tb}{2 \tb}
  (\de Z_{\Sbote  \Sbote}  + 
        \de Z_{\Sbotz \Sbotz}) \\
& \quad \non 
  + \frac{\mu + \Ab \tb}{\mb \tb}\Bigg[
 -i \frac{\MW \sin \theta_W}{e \tb}
  \Bigl(\La(0, \msbe^2,\msbe^2) + \La(0, \msbz^2, \msbz^2)\Bigr) \\[1.5mm]
& \qquad \non
  + \frac{\mb(\mu + \Ab  \tb)}{2 \tb} 
   (\de Z_{\Sbote  \Sbote}  + \de Z_{\Sbotz \Sbotz}) 
  +  (\msbe^2 - \msbz^2) \de \tsb \\[1.5mm] 
& \qquad  \non 
  - \tan \tsb \Bigl( \de  \msbz^2 -\cos^2 \tst \de \mste^2
  - \sin^2 \tst \de \mstz^2 + \sin 2 \tst (\mste^2 - \mstz^2)\de \tst
\\[1.5mm]& \qquad 
      + 2 \mt \de \mt
 \Bigr)
 \Bigg] 
\KKL \mu \bigl(\tb +
 \frac{1}{\tb}\bigr) 
 +2 \mb \tan \tsb \KKR^{-1} ,
  \label{dAbUIFartig}
\end{align} 
where the $Z$~factors are defined as
\begin{align}
\de Z_{\tilde{b}_i  \tilde{b}_i} &=
 -\frac{\Si_{\tilde{b}_{ii}}(\msbe^2) - 
        \Si_{\tilde{b}_{ii}}(\msbz^2)}
       {\msbe^2 -\msbz^2} \;. 
\end{align}
\end{itemize}
Again, 
the dependent counterterm for the $b$-quark mass is determined by
\refeq{deltaSbot} and
the constraint~\refeq{ms1CT}, but now inserting the above
specification~\refeq{dAbUIFartig} for $\de \Ab$, yielding
\begin{align} \non
\de \mb &= - \Bigl[
 - i \frac{\MW \sin \theta_W}{e\,\tb}
  \Bigl(\La(0, \msbe^2, \msbe^2) + \La(0, \msbz^2, \msbz^2) \Bigr) 
  + \frac{\mb(\mu + \Ab \tb)}{2 \tb} 
  (\de Z_{\Sbote  \Sbote}  + \de Z_{\Sbotz \Sbotz}) 
\\[1.5mm]
& \quad\ \non
+    (\msbe^2 - \msbz^2) \de \tsb  
+   \tan \tsb
  \Bigl( 
 - \sin 2 \tst (\mste^2 -\mstz^2)\de \tst 
  -  \de \msbz^2  
  + \cos^2 \tst \de \mste^2   
\\[1.5mm]
& \quad\ 
  + \sin^2 \tst \de \mstz^2    
  - 2  \mt \de \mt 
\Bigr) 
\Bigr]
\Bigl[
  \mu\bigl(\tb + \frac{1}{\tb}\bigr) 
  + 2 \mb \tan \tsb \Bigr]^{-1} ~.
\label{dmbmitthetauAbonshell}
\end{align}
The numerical value of $\mb$ in this scheme is obtained from 
\refeq{eq:mbdrbar2} and the (finite) difference of
the counterterms
given in \refeq{dmbmitthetauAbonshell} and \refeq{eq:dembdrbar}.

With the specification of $\de \mb$ in \refeq{dmbmitthetauAbonshell},
also the $\tilde{b}_1$-mass counterterm $\de \msbe^2$ in the general
relation~\refeq{ms1CT} is fully determined.


\subsection{Resummation in the \boldmath{$b/\Sbot$} sector}
\label{subsec:botresum}

The relation between the bottom-quark mass and the Yukawa coupling $h_b$,
which in lowest order reads $\mb = h_b v_1/\sqrt{2}$, receives radiative
corrections proportional to $h_b v_2 = h_b \tb \, v_1$. Thus, large
$\tb$-enhanced contributions can occur, which need to be properly taken
into account. As shown in \citeres{deltamb1,deltamb} the leading terms of 
\order{\alb(\als\tb)^n} can be resummed by using an appropriate
effective bottom Yukawa coupling. 

Accordingly, an effective bottom-quark mass is obtained by extracting
the UV-finite $\tb$-enhanced term $\De \mb$ from
\refeq{eq:mbdrbar2} (which enters through $\Si_{b_S}$) and writing it
as $1/(1 + \De \mb)$ into the  
denominator. In this way the leading powers of $(\als\tb)^n$ are
correctly resummed~\cite{deltamb1,deltamb}. This yields
\begin{equation}
\label{eq:mbdrbarresum}
\mb^{\drbarm, \text{MSSM}}(\mu^{\drbarm}) = 
  \frac{\mb^{\msbarm}(M_Z) b^{\text{shift}}  + \edz  \mb
  \Bigl(\Si^{\rm fin}_{b_L}({\mb}^2) + \Si^{\rm fin}_{{b}_R} ({\mb}^2)
\Bigr)
+ \mb\, \widetilde{\Si}^{\rm fin}_{b_S}(\mb^2)}{1 + \De \mb}~,
\end{equation}
where $\widetilde{\Si}_{b_S} \equiv \Si_{b_S} + \De \mb$ denotes the 
non-enhanced remainder of the scalar $b$-quark self-energy at
\order{\als}, and $b^{\text{shift}}$ is given in \eqref{dregdredbmassshift}.
The $\tb$-enhanced scalar part of the
$b$-quark self-energy, $\De \mb$, is given at \order{\als} by%
\footnote{
There are also corrections of \order{\alt} to $\De\mb$ that can be
resummed~\cite{deltamb}. These effects usually amount up to 5--10\% of
the \order{\als} corrections. Since in this paper we are interested only in the
\order{\alb\als} contributions to the MSSM Higgs sector, these
corrections have been neglected. Further corrections from subleading
resummation terms can be found in \citere{deltamb3}.
}%
\begin{align}
\label{eq:deltamb}
\De \mb =   \frac{2}{3 \pi} \als\tb\, \mu\, \mgl\,
 I (\msbe^2, \msbz^2, \mgl^2) ,
\end{align}
with
\begin{align}
I (\msbe^2, \msbz^2, \mgl^2) =
- \frac{\msbe^2 \msbz^2 \log(\msbz^2/\msbe^2) +
        \msbe^2 \mgl^2  \log(\msbe^2/\mgl^2) +
        \mgl^2 \msbz^2  \log(\mgl^2/\msbz^2)}
       {(\msbe^2 - \mgl^2) (\mgl^2 - \msbz^2) (\msbz^2 - \msbe^2)}~,
\label{eq:I}
\end{align}
and $\De \mb > 0$ for $\mu > 0$.

In the ``$\mb$ \drbar'' scheme we
use the effective bottom-quark mass as given in \refeq{eq:mbdrbarresum}
everywhere instead of the \drbar\ bottom quark mass 
(in particular, we use this bottom mass in the
sbottom-mass matrix squared, \refeq{Sfermionmassenmatrix}, from which
the sbottom mass eigenvalues are determined).
The numerical values of the bottom-quark mass in the other
renormalisation schemes can be obtained from \refeq{eq:mbdrbarresum} as
explained above, and from \refeq{mbdef} below.

We incorporate the effective bottom-quark mass of \refeq{eq:mbdrbarresum}
(or the correspondingly shifted value in the other renormalisation
schemes) into our one-loop results for the 
renormalised Higgs boson self-energies, which determine the Higgs boson masses
at one-loop order according to
\refeq{eq:proppole}--\refeq{renSE}. 
In this way the leading effects
of \order{\alb\als} are absorbed into the one-loop result. 
We refer to the genuine two-loop contributions,
which go beyond this improved one-loop result, as ``subleading
\order{\alb\als} corrections'' in the following.


\section{Numerical results}
\label{sec:numres}

\subsection{Evaluation}

If not mentioned explicitly in the text the default set of parameters shown in
\refta{tab:inputparameter} is used.
Large values of $\tan \beta$ and $|\mu|$ are chosen in order to
illustrate possibly large effects in the $b/\Sbot$ sector.

\begin{table}[!htb]
\renewcommand{\arraystretch}{1.7}
\begin{tabularx}{15.5cm}{|X|X|}\hline
\multicolumn{2}{|l|}{SM parameters:}\\  
\multicolumn{2}{|l|}{$\mt = 174.3 \gev$, 
      $\mb^{\msbarm}(M_Z) = 2.94 \gev$,}\\
\multicolumn{2}{|l|}{$\MZ = 91.1875 \gev$, 
$\MW = 80.426 \gev$, $\gf = 1.16639 \; 10^{-5}$}\\
\hline\hline
\multicolumn{2}{|l|}{parameters of the Higgs sector:}\\
\multicolumn{2}{|c|}{$\MA = 120 \gev$  \hspace{2.5cm} $\tb = 50$
  \hspace{2.5cm} $\mu = -1000 \gev$}\\ \hline\hline 
\multicolumn{2}{|l|}{soft-breaking parameters:}\\\hline 
for the gauginos:  & for the
 sfermions: 
 \\
  \multicolumn{1}{|c|}{$M_1 = \displaystyle{\frac{5}{3} \frac{\sin^2
   \theta_W}{\cos^2 \theta_W}} M_2$}  &  \multicolumn{1}{c|}{$M_L =
 M_{L_{\{\tilde{q}_i,\,\tilde{l}_i\}}} = 1000 \gev$ with $i = 1,\, 2,\,3$}
 \\
  \multicolumn{1}{|c|}{$M_2 = 100 \gev$} &
  \multicolumn{1}{c|}{$M_{\tilde{f}_R} = 1000 \gev$ with $f =
  u,\,c,\,t,\,d,\,s,\,b,\,e,\,\mu,\,\tau$} 
 \\
 \multicolumn{1}{|c|}{$M_3 = 1000 \gev$}  & \multicolumn{1}{c|}{$A_{\{u,\,c,\,t\}} =
  A_{\{d,\,s,\,b\}} = A_{\{e,\,\mu,\,\tau\}} = 2000 \gev$} \\\hline 
\end{tabularx}
\caption {\small Set of default input parameters.}
\label{tab:inputparameter}
\end{table}

We will mostly discuss the case of negative $\mu$, since
according to \refeqs{eq:mbdrbarresum}--(\ref{eq:I}) this sign of $\mu$ 
leads to a negative $\De\mb$ and therefore to an increase of the effective 
bottom-quark mass. This gives rise to an
enhancement of the corrections from the $b/\Sbot$ sector, see
\reffi{fig:deltamh}. While the negative sign of $\mu$ is disfavoured
from the comparison of the MSSM prediction~\cite{g-2MSSMf1l,g-2review}
with the experimental data on
the anomalous magnetic moment of the muon~\cite{g-2exp}, 
it would seem premature at this 
stage to completely disregard this possibility. For $\mu > 0$, on the
other hand, the corrections to the Higgs-boson masses from the $b/\Sbot$ 
sector will normally
not exceed the GeV level if the result is expressed in terms of an
appropriately chosen running bottom-quark mass (see \reffi{fig:deltamh}).
It should be noted, however, that the prospective experimental accuracy
on $\Mh$ at the LHC and the ILC will be significantly below the GeV
level, so that the inclusion of non-enhanced two-loop corrections will
be necessary in order to achieve the same level of precision for the
theoretical prediction (see the discussion below).

For the calculation of the Higgs boson masses presented below 
the complete \onel\ self-energies have been used, with 
$\tb$ renormalised in the \drbar\ 
scheme~\cite{dissMF,renormA,renormB} 
and with the $Z$~boson mass on-shell. 
At the \twol\ level, besides the \order{\alb\als}
corrections also the contributions \order{\alt\als} using
the one-loop sub-renormalisation 
of \refse{subsec:stoprenorm} have been included.
For simplicity we have neglected the \order{\alt^2} terms~\cite{mhiggsEP2}.
For the \order{\alt\als} corrections the top pole mass, 
$\mt = 174.3 \gev$, has been used. 
The inclusion of all known corrections and the new experimental top
quark mass value of $\mt = 178.0 \gev$~\cite{mtopexpnew} in our analysis
would yield an 
increase in $\Mh$ of \order{8 \gev}~\cite{mhiggsAEC}.
Therefore the mass values given
in our numerical analysis should not be viewed as predictions of
$\Mh$; they are rather illustrations of the $\als$-corrections
to the bottom Yukawa contributions at the two-loop level. 
(It should be noted that the chosen parameters are such that they are 
not in conflict with the experimental lower bounds on
$\Mh$~\cite{LEPHiggsSM,LEPHiggsMSSM}.)


\subsection{Comparison of the different renormalisation schemes}
\label{subsec:numanal}

\definecolor{lightblue}{cmyk}{1,0,0,0}
\definecolor{Blue}{rgb}{0,0,1}
\definecolor{Red}{named}{Red}
\definecolor{Green}{rgb}{0,0.9,0.2}
\definecolor{Black}{named}{Black}
\definecolor{Magenta}{named}{Magenta}
\definecolor{Royal}{named}{RoyalBlue}
\definecolor{Orange}{named}{Orange}
\definecolor{Purple}{named}{Purple}
\definecolor{Mahogany}{named}{Mahogany}
\definecolor{Brown}{named}{Brown}

\psfrag{MHH [GeV]}{{$\MH$ [GeV]}}
\psfrag{Mh0 [GeV]}{{$\Mh$ [GeV]}}
\psfrag{Delta Mh0 [GeV]}{{$\De\Mh$ [GeV]}}
\psfrag{MA0 [GeV]}{ $\MA$ [GeV]}
\psfrag{tan beta}{\raisebox{0.ex}{{$\tb$}}}
\psfrag{MUE [GeV]}{$\mu$ [GeV]}
\psfrag{MGl [GeV]}{$\mgl$ [GeV]}
\psfrag{MUE = -1000 GeV}{$\mu = -1000$ GeV}
\psfrag{MUE = 1000 GeV}{$\mu = 1000$ GeV}
\psfrag{MA0 = 120 GeV}{$\MA = 120$ GeV} 
\psfrag{MA0 = 700 GeV}{$\MA = 700$ GeV} 
\psfrag{MGl = 1000 GeV}{$\mgl =  1000$ GeV}
\psfrag{MGl = 1500 GeV}{$\mgl =  1500$ GeV}
\psfrag{MUE = -1000 GeV, MA0 = 120 GeV, MGl = 1000 GeV}
       {$\mu = -1000$ GeV, $\MA = 120$ GeV, $\mgl = 1000$ GeV}
\psfrag{MUE = 1000 GeV, MA0 = 120 GeV, MGl = 1000 GeV}
       {$\mu = 1000$ GeV, $\MA = 120$ GeV, $\mgl = 1000$ GeV}  
\psfrag{MUE = 1000 GeV, MA0 = 700 GeV, MGl = 1000 GeV}
       {$\mu = 1000$ GeV, $\MA = 700$ GeV, $\mgl = 1000$ GeV}  
\psfrag{MUE = -1000 GeV, TB = 50, MGl = 1000 GeV}
       {$\mu = -1000$ GeV, $\tb = 50$, $\mgl = 1000$ GeV}  
\psfrag{TB = 50}{ $\tan \beta = 50$}
\psfrag{TB = 50 GeV}{ $\tan \beta = 50$}
\psfrag{TB = 50, MA0 = 120 GeV, MGl = 1000 GeV}
       {$\tb = 50$, $\MA = 120$ GeV, $\mgl = 1000$ GeV}  
\psfrag{TB = 50, MA0 = 700 GeV, MGl = 1000 GeV}
       {$\tb = 50$, $\MA = 700$ GeV, $\mgl = 1000$ GeV}  

\psfrag{O(a_s a_t) with 1234567}
       {\order{\alt\als} with $\mb^{\drbarm,\text{MSSM}}$}
\psfrag{scheme1}
       {\hspace*{-3.4cm}\color{lightblue}{\order{\alb\als} : scheme: $\mb$ OS}}
\psfrag{scheme2}
       {\hspace*{-3.4cm}\Red{\order{\alb\als} : scheme: $\mb\,\drbarm$}}
\psfrag{scheme3}
       {\hspace*{-3.4cm}\Blue{\order{\alb\als} : scheme: $\Ab,\;\tsb \drbarm$}}
\psfrag{scheme4}
       {\hspace*{-3.4cm}\Green{\order{\alb\als} : scheme: $\Ab,\;\tsb$ OS}}

\psfrag{scheme1b}
       {\hspace*{-0.0cm}{\order{\alt\als} ($\mb^{\drbarm,\text{MSSM}}$)}}
\psfrag{scheme2b}
       {\hspace*{-0.0cm}\Red{\order{\alb\als} : $\mb\,\drbarm$}}
\psfrag{scheme3b}
       {\hspace*{-0.0cm}\Blue{\order{\alb\als} : $\Ab,\;\tsb~\drbarm$}}
\psfrag{scheme4b}
       {\hspace*{-0.0cm}\Green{\order{\alb\als} : $\Ab,\;\tsb$ OS}}
\psfrag{scheme2b1L}
       {\hspace*{-0.0cm}{\order{\alt\als} : $\mb\,\drbarm$ for \order{\alb}}}
\psfrag{scheme3b1L}
       {\hspace*{-0.0cm}{\order{\alt\als} : $\Ab,\;\tsb~\drbarm$ for \order{\alb}}}
\psfrag{scheme4b1L}
       {\hspace*{-0.0cm}{\order{\alt\als} : $\Ab,\;\tsb$ OS for \order{\alb}}}

\psfrag{O(a_s a_t) default param.}{}
\psfrag{UIF-type scheme O(a_s a_t)}
       {\hspace*{-0.0cm}\order{\alt\als} : $\Ab,\;\tsb$ OS for \order{\alb}}
\psfrag{UIF-type scheme} 
       {\hspace*{-0.3cm}\order{\alb\als} : $\Ab,\;\tsb$ OS}
\psfrag{with Pietros code O(a_s a_t)}
      {\hspace*{-0.4cm}\order{\alt\als} : $\msb,\;\Ab$~OS for \order{\alb}}
\psfrag{with Pietros code}  
       {\hspace*{-0.7cm}\order{\alb\als} : $\msb,\;\Ab$~OS}

\psfrag{diffscheme2mehrplatz}
       {\hspace*{+1.0cm}{\order{\alb\als} : $\mb\,\drbarm$}}
\psfrag{diffscheme3}
       {\hspace*{-1.5cm}{\order{\alb\als} : $\Ab,\;\tsb~\drbarm$}}
\psfrag{diffscheme4}
       {\hspace*{-1.5cm}{\order{\alb\als} : $\Ab,\;\tsb$ OS}}

In order to compare the different renormalisation schemes, the
parameters entering the one-loop result have to be transformed according
to the different renormalisation prescriptions. As our default for which
the input parameters are fixed we have chosen the ``$\mb$ \drbar'' scheme, 
where $\mb$ and $\Ab$ are defined as \drbar\ parameters.
As explained in \refse{sec:ren}, the parameters are converted to a
different renormalisation scheme RS 
(with counterterms \nolinebreak $\de x^{\text{RS}}$)
with the help of the following transformations, 
\begin{align}
\label{mbdef}
\mb^{\text{RS}} &= 
\mb^{\drbarm} -\de\mb^{\text{RS}}|_{\text{finite}}\, , \\ 
\label{Abdef}
\Ab^{\text{RS}} &= 
\Ab^{\drbarm} -\de \Ab^{\text{RS}}|_{\text{finite}}\;,
\end{align}
and analogously for the other parameters.
If not stated differently, the
\drbar\ scale has always been chosen as $\mu^{\drbarm} = \mt$. 
As an example, 
in \refta{wertetanbeta30} and \refta{wertetanbeta50} numerical values
for the bottom quark mass, $\Ab$ 
and the sbottom masses in the different schemes 
(see \refta{tab:sbotren}), are
shown for $\tb = 30$ and $\tb = 50$ and using the default
values given in \refta{tab:inputparameter} otherwise.

\begin{table}[htb!]
\renewcommand{\arraystretch}{1.4}
\BC
\begin{tabular}{|c||r|r|r|}
 \hline
scheme & $\mb$ [GeV] & $\Ab$ [GeV] & $\msbe$ [GeV] \\\hline\hline
$\mb$ \drbar         & 3.79 & 2000.00  & 1059.95 \\ \hline
$\Ab$, $\tsb$ \drbar & 3.04 & 2000.00  & 1039.50 \\ \hline
$\Ab$, $\tsb$ OS     & 2.99 & 2332.81  & 1039.04 \\ \hline
$\mb$ OS             & 3.77 & -4284.56 & 1039.25 \\ \hline
\end{tabular}
\EC
\caption{Values of the bottom quark mass, $\Ab$ and $\msbe$
  in the different schemes for $\tb = 30$ and $\mu = -1000 \gev$. The
  value of $\msbz$, which is 
  renormalised on-shell (see \refeq{eq:msbz}), is the same in all four 
  schemes, $\msbz = 938.44$~GeV.
\label{wertetanbeta30}} 
\end{table}

\begin{table}[htb!]
\renewcommand{\arraystretch}{1.4}
\BC
\begin{tabular}{|c||r|r|r|}
\hline
scheme & $\mb$ [GeV] & $\Ab$  [GeV]& $\msbe$ [GeV] \\\hline\hline
$\mb$ \drbar         & 5.82 & 2000.00 & 1142.16 \\\hline
$\Ab$, $\tsb$ \drbar & 5.26 & 2000.00 & 1117.93 \\\hline
$\Ab$, $\tsb$ OS     & 5.24 & 2219.40 & 1118.02 \\\hline
$\mb$ OS             & 4.93 & 6508.12 & 1122.04 \\\hline
\end{tabular}
\EC
\caption{Values of the bottom quark mass, $\Ab$ and $\msbe$
  in the different schemes for $\tb = 50$ and $\mu = -1000 \gev$. The
  value of $\msbz$ 
  is the same in all four schemes, $\msbz = 836.48$~GeV.
\label{wertetanbeta50}}
\end{table}

The values given in \refta{wertetanbeta30} and \refta{wertetanbeta50}
indicate that the ``$\mb$ OS'' scheme leads to huge corrections in $\Ab$
that invalidate the applicability of this scheme. The other schemes give
rise to only moderate shifts in the parameters.

The reason for the problematic behaviour of the ``$\mb$ OS'' scheme is
easy to understand. The renormalisation condition in the ``$\mb$ OS''
scheme is a condition on the sbottom mixing angle $\tsb$ and thus on the 
combination $(\Ab - \mu \tb)$ (see
\refeq{deltaSbot}). In parameter regions where
$\mu \tb$ is much larger than $\Ab$, the counterterm $\de\Ab$ receives
a very large finite shift when calculated from the counterterm 
$\de\tsb$.
More specifically, $\de\Ab$ as given in \refeq{Abparameter} contains
the contribution
\BEA
\label{largedAb}
\de\Ab &=& \frac{1}{\mb} \left[- \frac{\de\mb}{2\,\mb} \, 
             (\msbe^2 - \msbz^2) \sin 2\tsb + \ldots \right] \non \\
       &=& \frac{1}{\mb} \left[- \de\mb (\Ab - \mu \, \tb) + \ldots
           \right] , 
\EEA
that can give rise to very large corrections to $\Ab$. This problem is
avoided in the other renormalisation schemes introduced in 
\refta{tab:sbotren}, where the renormalisation condition is applied
directly to $\Ab$, rather than deriving $\de\Ab$ from the
renormalisation of the mixing angle.

\smallskip
We now turn to the numerical comparison of the different
renormalisation schemes.
As discussed above, the $\tb$-enhanced terms of \order{\alb\als} entering
via $\De \mb$ have been absorbed into the one-loop result.
The meaning of the various curves in the following figures 
is specified as (see also \refta{tab:sbotren}):
\begin{itemize}
\item
dashes with dots (black): \order{\alt\als} with $\mb^{\drbarm,\text{MSSM}}$,
results without subleading two-loop \order{\alb\als}\ terms
\item
dot-dash (light blue): 
``$\mb$ OS'' scheme for
subleading \twol\ \order{\alb\als} terms
\item
solid (red): ``$\mb$ \drbar'' scheme for
subleading \twol\ \order{\alb\als} terms
\item
dotted (dark blue): ``$\Ab$, $\tsb$ \drbar'' scheme for
subleading \twol\ \order{\alb\als} terms
\item
dashes with stars (green): ``$\Ab$, $\tsb$ OS'' scheme for
subleading \twol\ \order{\alb\als} terms
\end{itemize}

\begin{figure}[htb!]
\begin{center}
\epsfig{figure=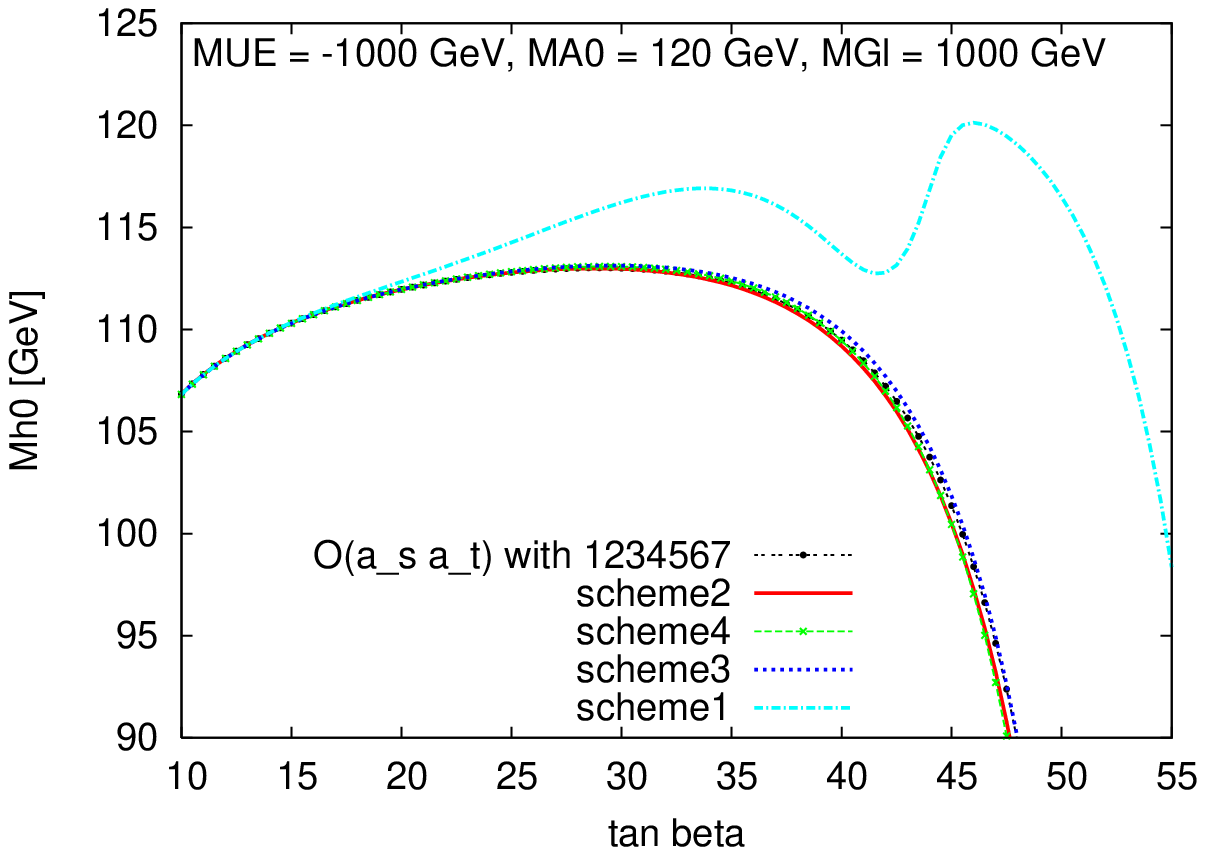, width=14cm,height=10cm}\\[1.5cm]
\epsfig{figure=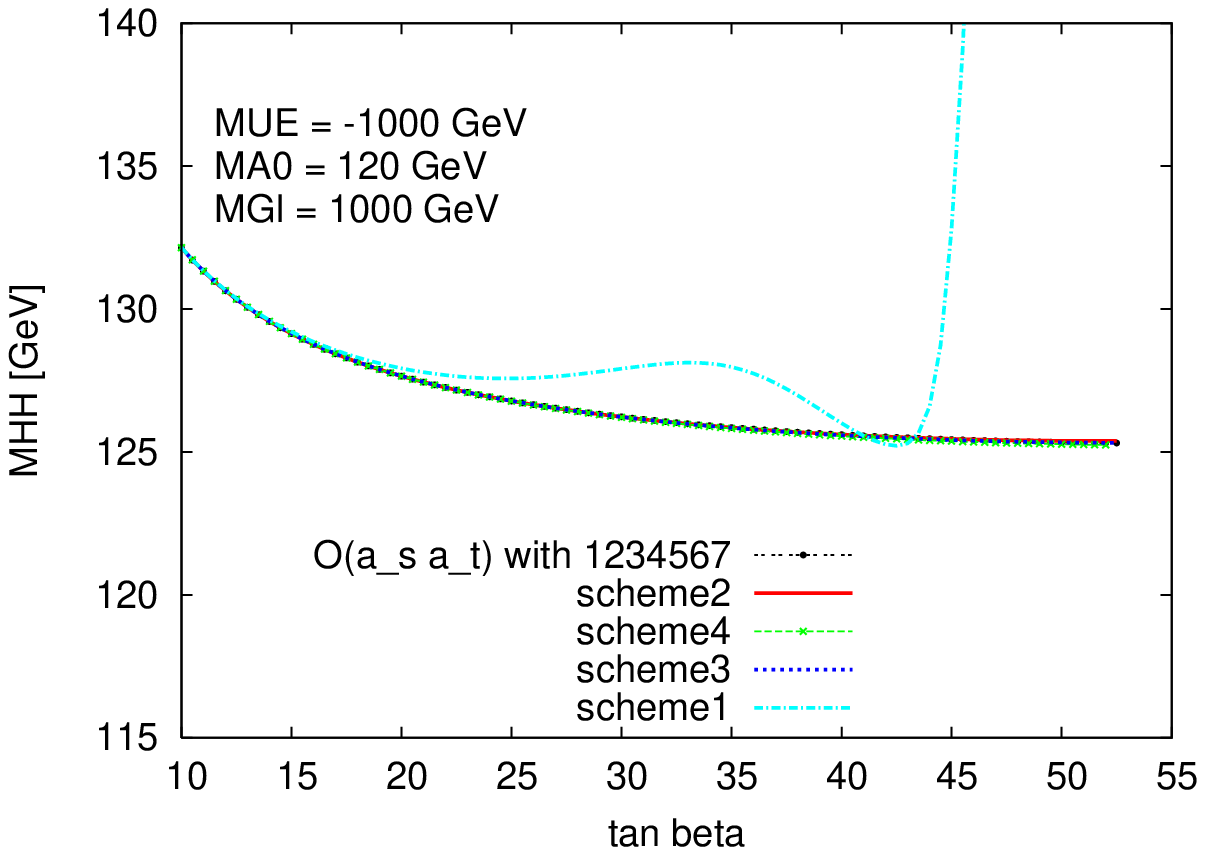, width=14cm,height=10cm}
\caption{
$\tb$-dependence of $\Mh$ and $\MH$ for $\MA = 120 \gev$ and $\mu$ negative.
}
\label{fig:mhtb_MAsmall_MUneg}
\end{center}
\end{figure}

We start our analysis of the different renormalisation schemes by
comparing the results for $\Mh$ and $\MH$ as a function of $\tb$ 
in \reffi{fig:mhtb_MAsmall_MUneg}.
The other parameters are as given in \refta{tab:inputparameter}.
As expected from the discussion of
\refta{wertetanbeta30} and \refta{wertetanbeta50}, the 
``$\mb$ OS'' scheme gives rise to artificially large corrections and 
shows very large deviations from the other
schemes for intermediate and large values of $\tb$. This 
behaviour is even more pronounced for $\MH$ than for $\Mh$, as 
can be seen in the lower plot of \reffi{fig:mhtb_MAsmall_MUneg}. 
These extremely large corrections are a consequence of the large 
contributions to the counterterm of
the parameter~$\Ab$ (see \refeq{largedAb}).
The Higgs self-energy contribution from virtual
sbottoms contains a term proportional to $\Ab^2$.
Using as input a value for $\Ab$ according to
\eqref{Abdef}, very large contributions proportional to
$(\de\Ab)^2$ are introduced. 
These corrections are more pronounced
in $\Si_{HH}$, where they enter like $(\Ca \Ab)^2$, than in
$\Si_{hh}$, where they enter like $(\Sa \Ab)^2$ ($|\al| \ll 1$ in our
analysis). The unacceptably large contributions to $\de\Ab$ in the
``$\mb$~OS'' scheme invalidate a perturbative treatment in this scheme.
We therefore discard this scheme in the following and focus our
discussion on the other three schemes defined in 
\refta{tab:sbotren}.

The other schemes all give similar and numerically well-behaved
results, where $\Mh$ starts to decrease
rapidly with $\tb$ for $\tb \gsim 40$. Negative mass squares
are reached at $\tb \simeq 53$. 
The main effect comes from the leading contributions of
\order{\alb\als} that enter via the resummation of $\De\mb$, see
\refeq{eq:mbdrbarresum}. 
The decrease with increasing $\tb$ is mainly due to the dependence of 
$\De\mb \sim \mu\tb$ in \refeq{eq:deltamb}. The subleading
\order{\alb\als} corrections, which arise from the genuine two-loop
diagrams, are of \order{1\gev}. The differences between the three
renormalisation schemes are of similar size. 
For this particular parameter choice the ``$\Ab$, $\tsb$ \drbar'' scheme 
enhances $\Mh$, whereas the other two schemes decrease~$\Mh$ compared to
the case where the genuine two-loop corrections are omitted.

\begin{figure}[htb!]
\begin{center}
\epsfig{figure=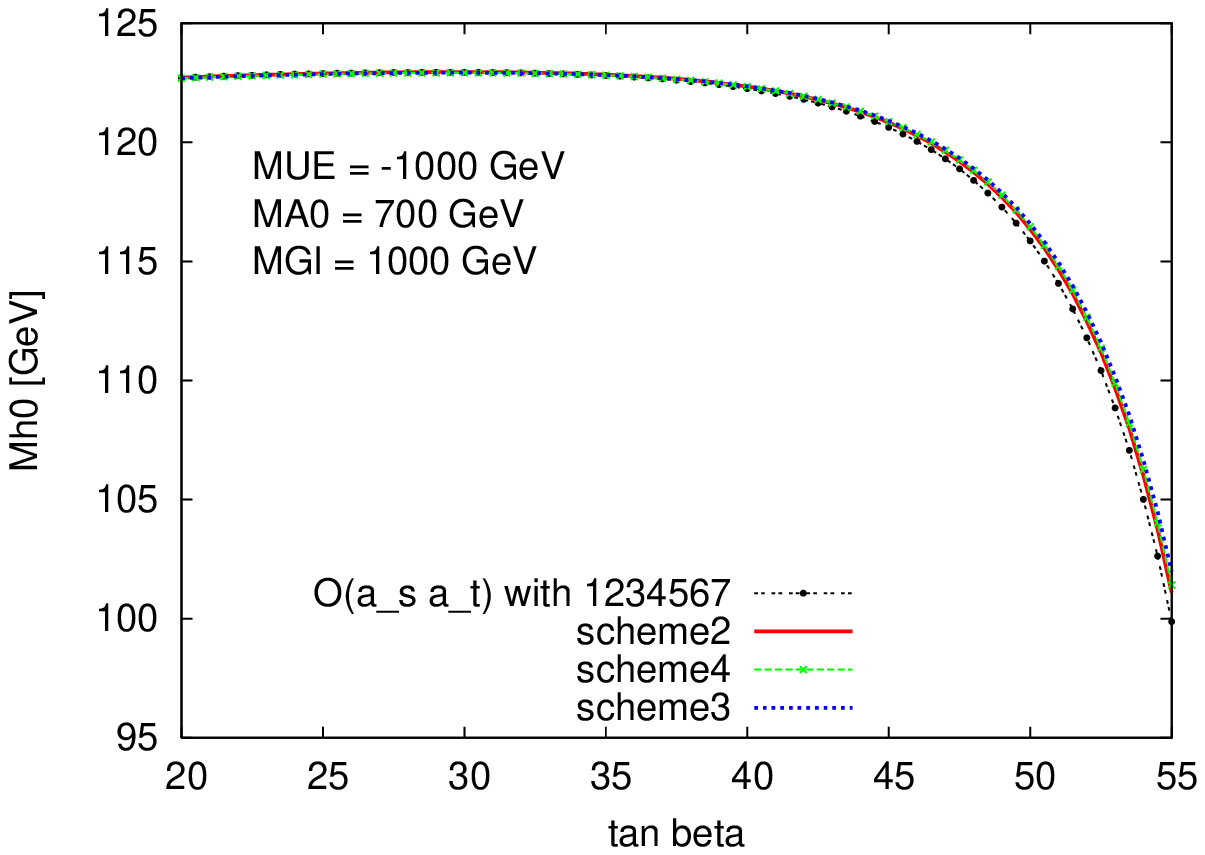, width=14cm,height=10cm}
\caption{
$\tb$-dependence of $\Mh$ for $\MA = 700 \gev$ and $\mu$ negative.
}
\label{fig:mhtb_MAlarge_MUneg}
\end{center}
\end{figure}

In \reffi{fig:mhtb_MAlarge_MUneg} we show $\Mh$ as a function of $\tb$
for the same parameters as in \reffi{fig:mhtb_MAsmall_MUneg}, but with
$\MA = 700 \gev$. This results in general in larger $\Mh$ values, but
the general behaviour as a function of $\tb$ is the same as for 
$\MA = 120 \gev$; $\Mh$ drops steeply for large $\tb$ values.
In all three schemes the subleading terms increase $\Mh$ by a few GeV,
depending on $\tb$.  

\begin{figure}[t!]
\begin{center}
\epsfig{figure=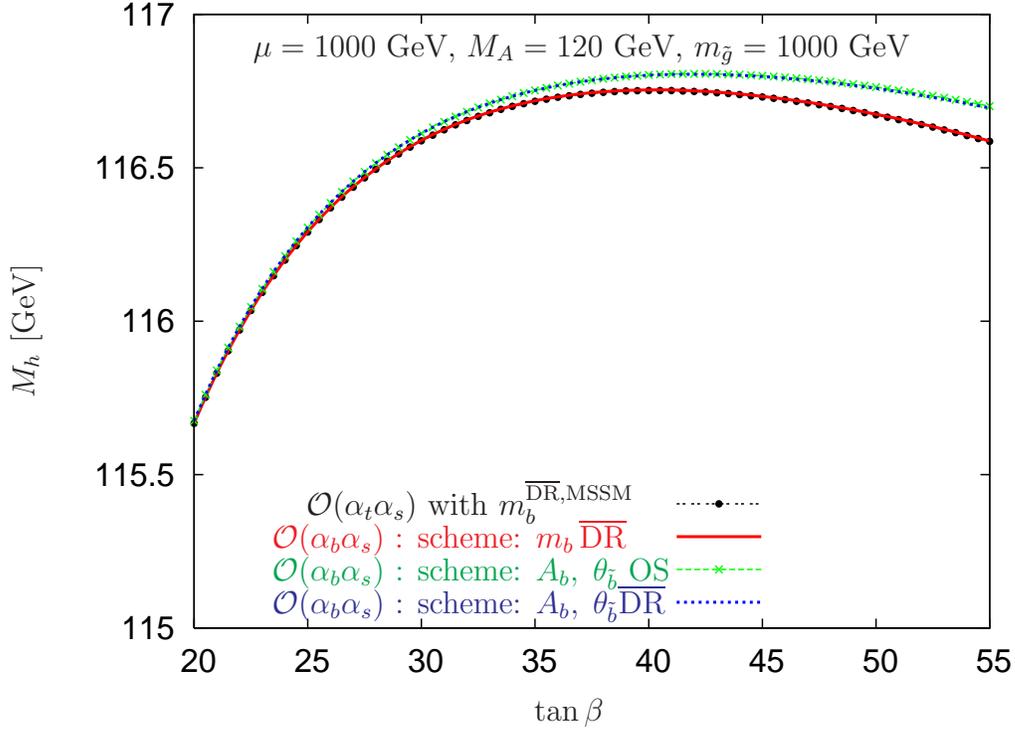, width=14cm,height=10cm}\\[1cm]
\epsfig{figure=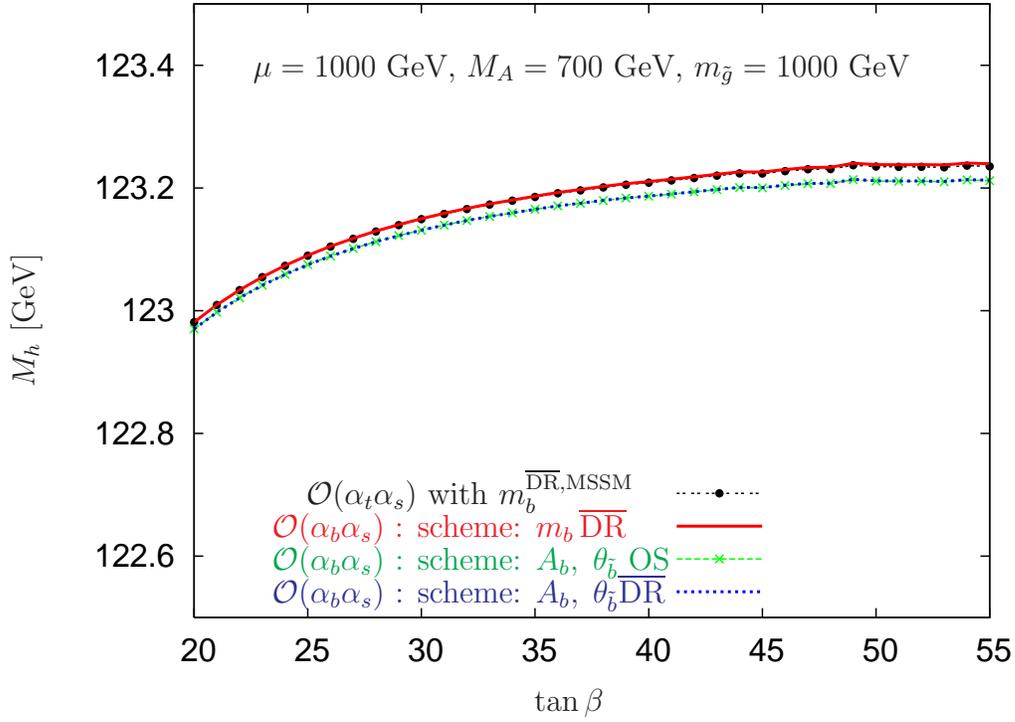, width=14cm,height=10cm}
\caption{
$\tb$-dependence of $\Mh$ for $\MA = 120 \gev$ (upper plot)
and $\MA = 700 \gev$ (lower plot) for positive $\mu$.
}
\label{fig:mhtb_MUpos}
\end{center}
\end{figure}

As discussed above, large corrections from the $b/\Sbot$ sector are
only expected for negative values of $\mu$.
In \reffi{fig:mhtb_MUpos} 
we show the results for $\Mh$ as a function of $\tb$ with positive $\mu$
and $\MA = 120$ and $700 \gev$, respectively. The other parameters are
given in \refta{tab:inputparameter}. 
The positive sign of $\mu$ results in a positive $\De\mb$ and thus a
smaller numerical value of $\mb^{\drbarm,\text{MSSM}}$.
As expected%
\footnote{
See also the discussion in \citere{mhiggsEP4}, where the
opposite sign convention for $\mu$ is used.
}%
, the variation of
$\Mh$ with $\tb$ is much smaller than for negative $\mu$. Both, the
leading corrections, i.e.\ the $\tb$ enhanced terms of
\order{\alb\als}, as well as
the subleading corrections are at the level of \order{100 \mev}. 
The ``$\mb$ \drbar'' scheme does not show any visible corrections
beyond the resummed contributions. 
This leads to the conclusion that for positive $\mu$ the corrections
beyond the \onel\ level coming from the $b/\Sbot$~sector are 
sufficiently well under control. However, in view of the fact that the
anticipated ILC accuracy on $\Mh$~\cite{tesla,orangebook,acfarep} and
the parametric uncertainty of the theory prediction from the ILC
measurement of the top-quark mass~\cite{tbexcl,deltamt} 
will both be about 100 MeV, ultimately the aim will be to
reduce the theoretical uncertainties from unknown higher-order
corrections to at least this level. This will require the inclusion of
all two-loop corrections (and a significant part of 
corrections beyond two-loop order).
For the further analysis in this paper we focus on negative values of
$\mu$.

\begin{figure}[t!]
\begin{center}
\epsfig{figure=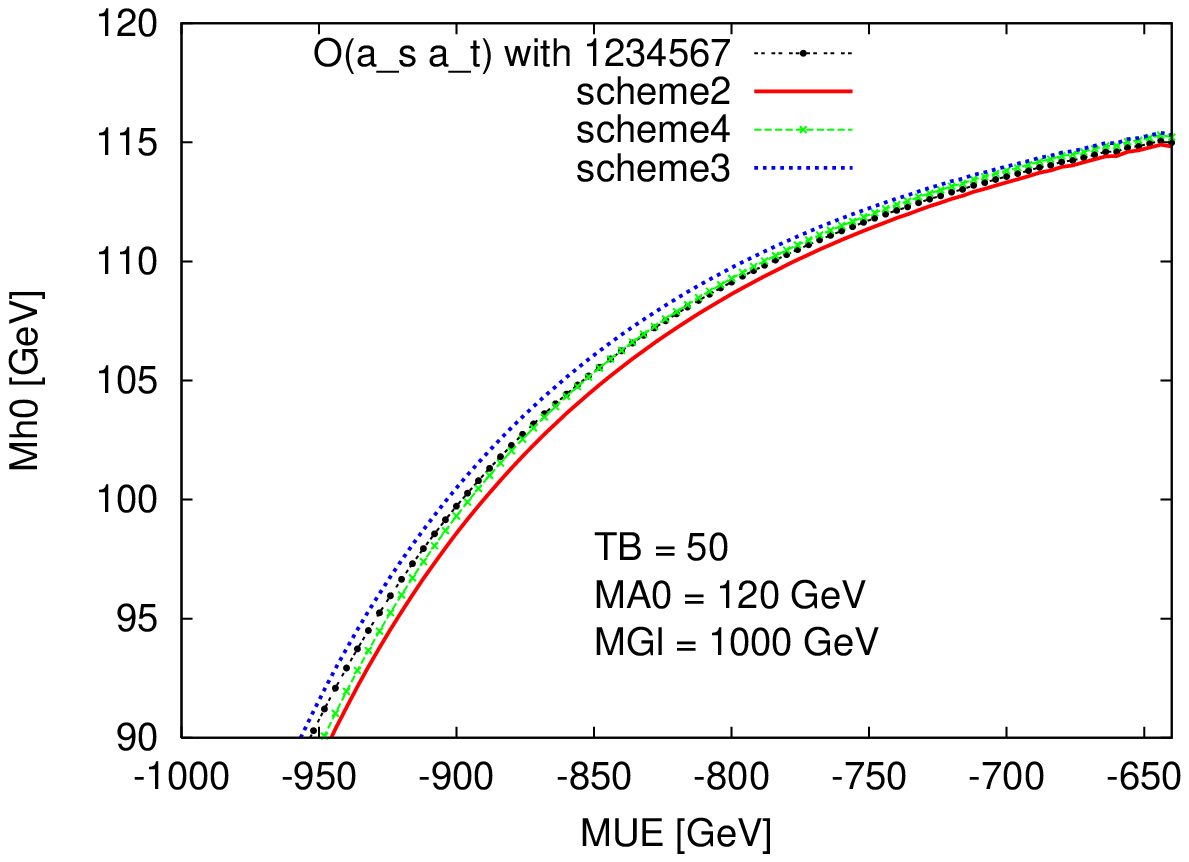, width=14cm,height=10cm}\\[1cm]
\epsfig{figure=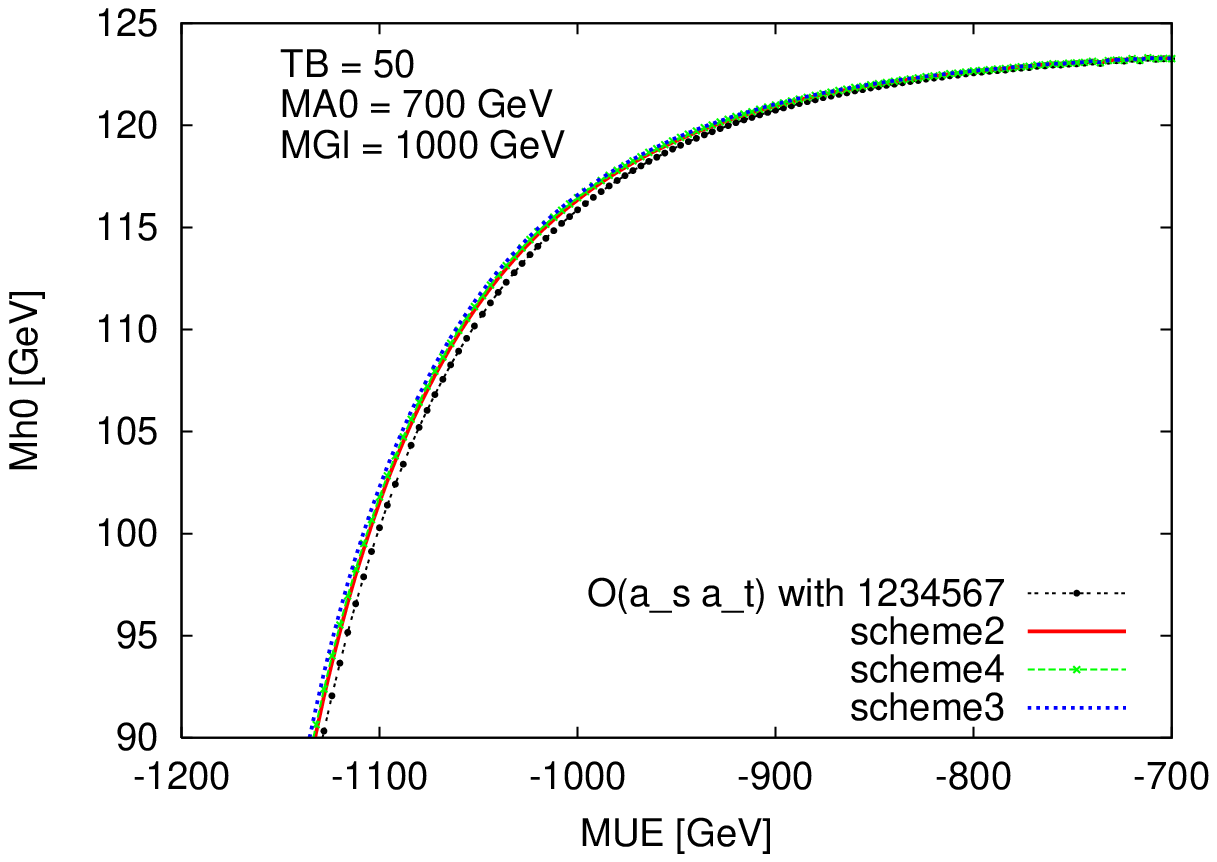, width=14cm,height=10cm}
\caption{
$\mu$-dependence of $\Mh$ for $\MA = 120 \gev$ (upper plot)
and $\MA = 700 \gev$ (lower plot) for $\tb = 50$.
}
\label{fig:mhmu}
\end{center}
\end{figure}

The variation of $\Mh$ with $\mu$ (for $\mu < 0$) for $\tb = 50$ 
is shown in \reffi{fig:mhmu}.
As can be expected from
\refeq{eq:deltamb} the corrections at \order{\alb\als} increase with
increasing~$|\mu|$. Typically the genuine two-loop contributions are  
of \order{1 \gev}.
For large $\MA$ all the
schemes lead to an increase of $\Mh$,
whereas for small $\MA$ both negative and positive shifts can occur.
Differences in the $\Mh$ predictions induced by the different
renormalisation schemes are below the GeV level for large $\MA$.

\begin{figure}[t!]
\begin{center}
\epsfig{figure=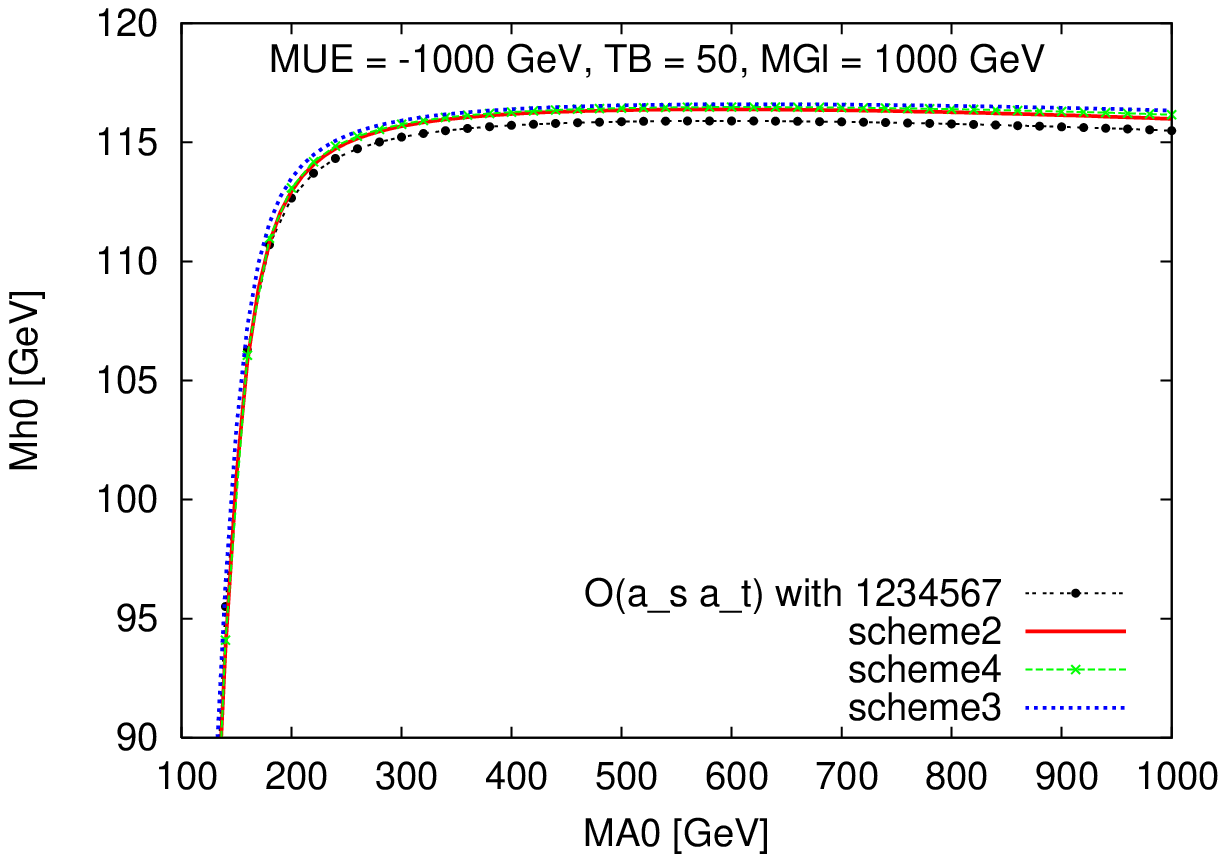, width=14cm,height=10cm}
\caption{
$\MA$-dependence of $\Mh$ for $\tb = 50$ and $\mu$ negative.
}
\label{fig:mhMA}
\end{center}
\end{figure}

In \reffi{fig:mhMA} the dependence of $\Mh$ on $\MA$ is shown for the
different renormalisation schemes, with the other 
default parameters from \refta{tab:inputparameter}. 
For $\MA \gsim 200 \gev$ the subleading terms of
all three schemes enhance $\Mh$ by \order{1 \gev}. A decrease only
occurs for small values of $\MA$, depending on the scheme.
The differences
in the $\Mh$ prediction resulting from the use of different
renormalisation schemes decrease for $\MA \gsim 200 \gev$ to
\order{0.1 \gev}.

%
\begin{figure}[htb!]
\begin{center}
\epsfig{figure=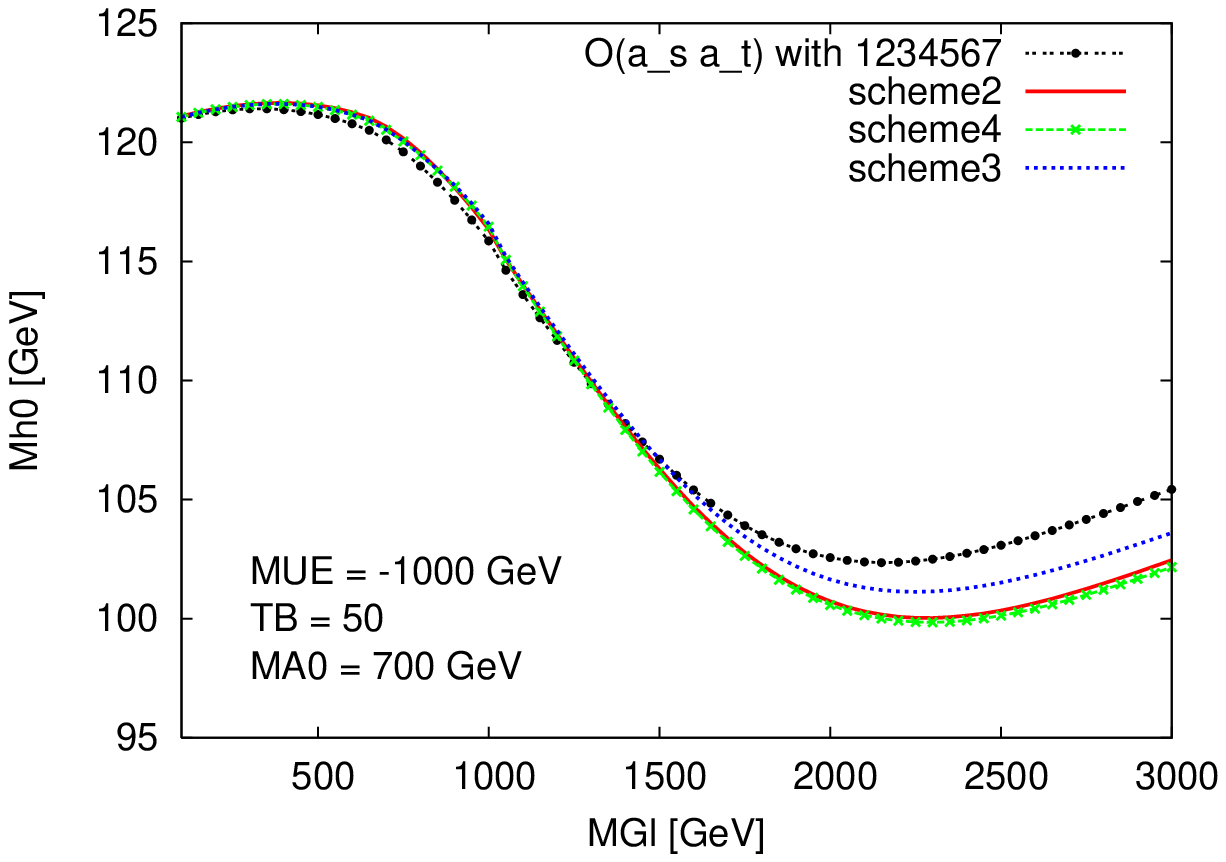, width=14cm,height=10cm}
\caption{
$\mgl$-dependence of $\Mh$ for $\MA = 700 \gev$, $\tb = 50$ and $\mu$
negative. 
}
\label{fig:mhmgl}
\end{center}
\end{figure}

In \reffi{fig:mhmgl} it can be seen that the behaviour of the corrections 
strongly depends on the choice of $\mgl$. The figure shows $\Mh$ as a
function of $\mgl$ for $\mu = -1000 \gev$, $\tb = 50$ and $\MA = 700 \gev$.
For $\mgl \lsim 1000 \gev$ all
schemes lead to an increase of $\Mh$ from the subleading
\order{\alb\als} corrections. For $\mgl \gsim 1500 \gev$, on the other
hand, all schemes lead to a decrease, where the size of the individual
corrections also strongly varies with $\mgl$. Accordingly, the relative
size of the corrections in the different schemes also
varies with $\mgl$. Corrections up to about $3 \gev$ are possible. 
The differences between the three schemes
are of \order{2 \gev} for large $\mgl$. 
It should be noted that the effects of the higher-order corrections to
$\Mh$ do not decouple with large $\mgl$. The corrections at
\order{\alt\als}~\cite{mhiggslong} as well as \order{\alb\als} grow
logarithmically in the renormalisation schemes that we have adopted.

The above analysis of the three schemes 
``$\mb$ \drbar'', ``$\Ab$, $\tsb$ \drbar'', and ``$\Ab$, $\tsb$ OS'' in
various parameter regions yields numerically well-behaved and physically
meaningful results. As there is no clear preference for one of the
schemes on physical grounds, the difference between the results obtained
in the three schemes can be interpreted as an indication of the possible
size of missing higher-order corrections. 
The
size of the individual corrections and also the differences
between the renormalisation schemes sensitively depend on the input
parameters. Typically we find that the genuine two-loop corrections 
in the $b/\Sbot$~sector yield a shift in $\Mh$ of \order{1 \gev}. 
\mbox{The differences between the three schemes are usually somewhat smaller.}


\subsection{Numerical analysis of the renormalisation scale dependence}
\label{subsec:numanalmudim}

\begin{figure}[hb!]
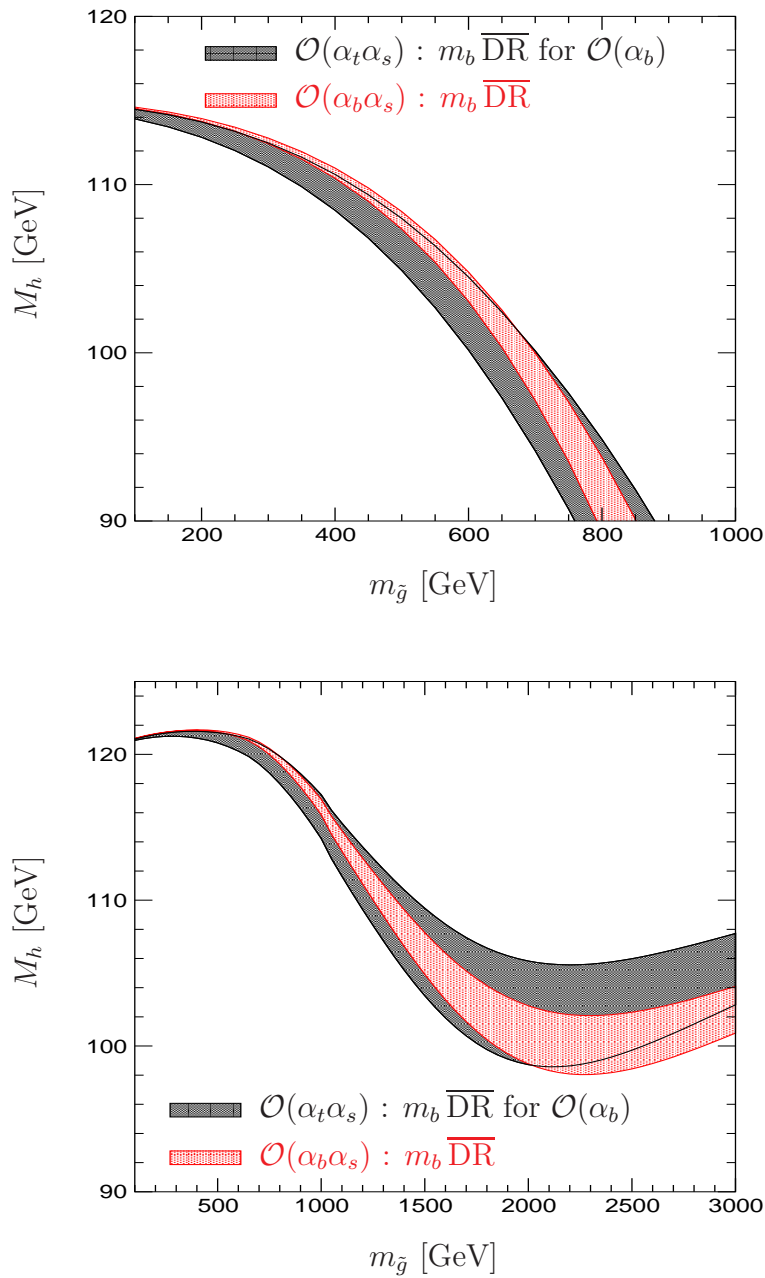

\begin{center}
\epsfig{figure=plots/MhmudimMA120tb50mun03.cl.eps,
                                          width=10cm,height=7.9cm}\\[1cm]
\epsfig{figure=plots/MhmudimMA700tb50mun03.cl.eps, width=10cm,height=7.9cm}
\caption{
$\mudim$ dependence of $\Mh$ as a function of $\mgl$ for
$\MA = 120 \gev$ (upper plot) and $\MA = 700 \gev$ (lower plot) for 
$\mu = -1000 \gev$, $\tb = 50$. The black area
corresponds to the \order{\alt\als} result including resummation,
i.e.\ the result without the subleading two-loop \order{\alb\als} terms.
}
\label{fig:mhmudim}
\end{center}
\end{figure}

While in the previous section we compared the results of different
renormalisation scheme, we now focus on the ``$\mb$~\drbar'' scheme
and investigate the effect of varying the renormalisation scale of the 
\order{\alb\als} result obtained in this scheme. We vary the scale
within the interval $\mt/2 \le \mudim \le 2\,\mt$, resulting in a shift
which is formally of \order{\alb\als^2}.
The results are shown as a function of $\mgl$ for $\tb = 50$ in
\reffi{fig:mhmudim} for $\MA = 120 \gev$ and $\MA = 700 \gev$.

The $\mudim$ variation of the leading contribution (the \order{\alt\als}
result including resummation) is shown as the
dark shaded (black) band. The results including the subleading
corrections in the ``$\mb$ \drbar'' scheme are shown as a light shaded
(red) band. It can be seen that the variation with $\mudim$ is
strongly reduced by the inclusion of the subleading contributions. 
The variation with $\mudim$ within the ``$\mb$ \drbar'' scheme is tiny
for $\mgl \lsim 500 \gev$, and reaches $\pm 2 \gev$  for large
$\mgl$ values. 
Thus, the $\mudim$ variation causes a similar shift in $\Mh$ as the
comparison between the three renormalisation schemes discussed above.

We have also analysed the variation with $\mudim$ in the case 
$\mu > 0$, which is not shown here. As for negative $\mu$, the
variation with $\mudim$ is of the same order as the differences
between the three renormalisation schemes, see
\reffi{fig:mhtb_MUpos}. 
Therefore, for $\mu > 0$ the unknown higher-order corrections to $\Mh$
from the $b/\Sbot$~sector can be estimated to be of \order{100 \mev}.


\subsection{Comparison with existing calculations}
\label{subsec:UIFcomp}

Finally we compare our result with the existing calculation of the
\order{\alb\als} corrections presented in \citere{mhiggsEP4}. 
The renormalisation employed there consists of an on-shell
renormalisation of the two scalar bottom masses and the on-shell
condition for $\Ab$ shown in \refse{subsubsec:AbtsbOS}. 
We denote it as ``$\msb,\;\Ab$~OS'' renormalisation.
Thus, the differences between our ``$\Ab,\;\tsb$~OS'' and
the ``$\msb,\;\Ab$~OS'' renormalisation are the different treatment of the
$\msbe$ renormalisation, 
as well as the treatment of $\tb$. We kept
$\tb$ as a free parameter, whereas in \citere{mhiggsEP4} it was set to
infinity in the subleading \order{\alb\als} corrections. 
In \citere{mhiggsEP4} the shift of the sbottom masses due to the
SU(2)-invariance was taken into
account in the numerical evaluation of the sbottom masses following the
prescription in \citere{bartl} (see also \citere{delrhosusy2loop}).

\begin{figure}[th!]
\begin{center}
\epsfig{figure=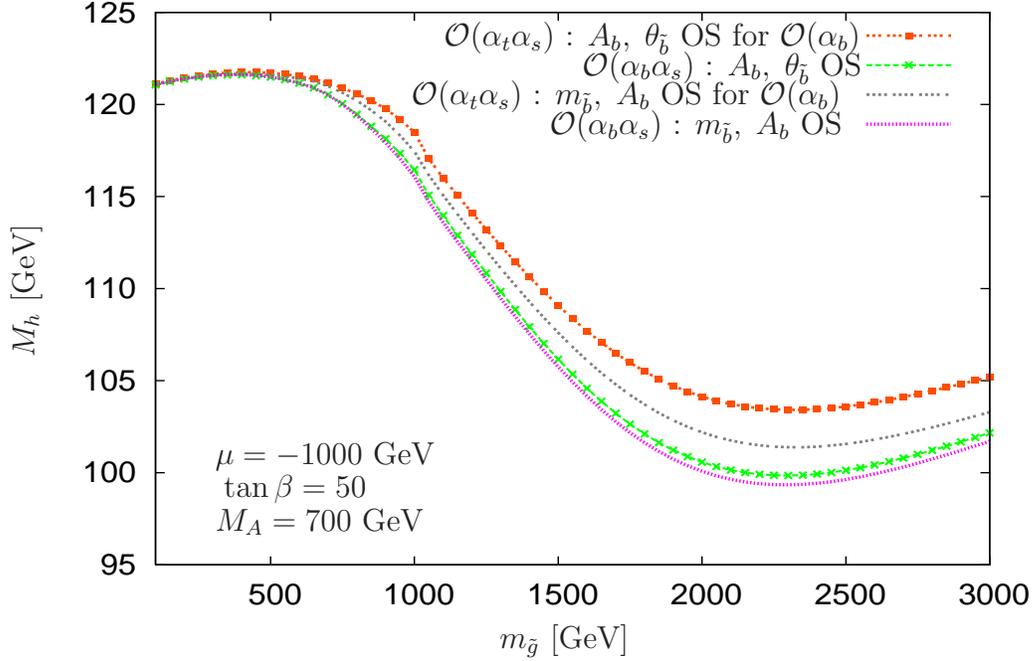, width=14cm,height=9.0cm}
\caption{
Comparison of our \order{\alb\als} result for $\Mh$
in the ``$\Ab,\;\tsb$~OS'' scheme and the
result of \citere{mhiggsEP4} (``$\msb,\;\Ab$~OS'' scheme) as a function
of $\mgl$.
The \order{\alt\als} results in the two schemes, where the subleading 
\order{\alb\als} corrections are omitted (using the appropriate renormalised
parameters), are also shown.
}
\vspace{-2em}
\label{fig:UIFcomp}
\end{center}  
\end{figure}

Our result for $\Mh$ in the ``$\Ab,\;\tsb$~OS'' scheme is compared with 
the result of \citere{mhiggsEP4} in \reffi{fig:UIFcomp}. For the 
implementation of the latter (``$\msb,\;\Ab$~OS'' scheme) 
the Fortran code of \citere{mhiggsEP4}
for the numerical evaluation of the $\mathcal O(\alpha_s \alpha_b)$ corrections 
to the Higgs-boson self-energies has been used~\cite{pietro}.
Thereby the input values were determined according to
\eqref{mbdef} and \eqref{Abdef}. Using these input values for $\Ab$ and
$\mb$ the sbottom masses were calculated taking the sbottom mass shift
into account~\cite{bartl}.
$\Mh$ is shown as function of $\mgl$ for 
$\mu < 0$, $\tb = 50$, and $\MA = 700 \gev$. Our result in the
``$\Ab,\;\tsb$~OS'' scheme is shown as the dash-star (green) curve, while
the result of \citere{mhiggsEP4} (``$\msb,\;\Ab$~OS'' scheme) is given by the 
fine-dotted (pink) curve. The leading contribution in the two schemes,
i.e.\ the \order{\alt\als} result including resummation, is also shown:
the light-dot-dashed (orange)
curve shows the \order{\alt\als} result using the 
``$\Ab,\;\tsb$~OS'' renormalised
parameters; the corresponding result for the ``$\msb,\;\Ab$~OS''
renormalised parameters is shown as the light-dotted (gray) curve.

\reffi{fig:UIFcomp} shows that the \order{\alt\als} results in the two
schemes differ from each other by up to $2 \gev$ for large $\mgl$. The
inclusion of the subleading two-loop corrections reduces this difference
significantly. Our result in the ``$\Ab,\;\tsb$~OS'' scheme agrees with the
result of \citere{mhiggsEP4} to better than $0.5 \gev$.

%
\begin{figure}[hb!]
\begin{center}
\epsfig{figure=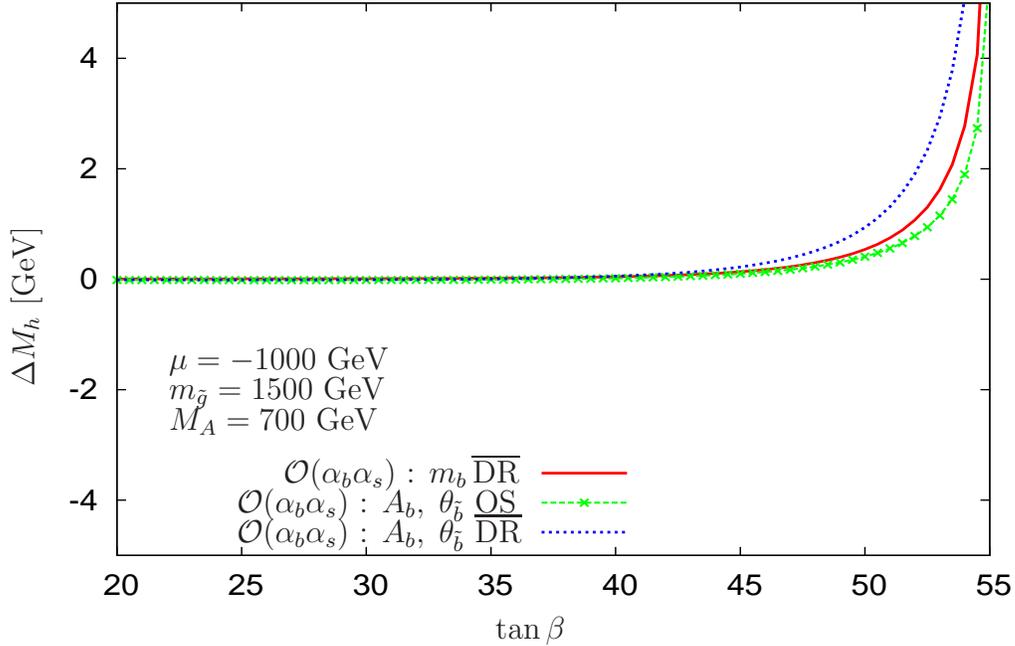, width=14cm,height=9.0cm}
\caption{
Comparison of our \order{\alb\als} result in three different 
renormalisation schemes and the result of \citere{mhiggsEP4}.
The three curves for $\De\Mh$ show the difference between our result in each
of the three schemes and the result of \citere{mhiggsEP4} 
as a function of $\tb$.
}
\label{fig:UIFcomp2}
\end{center}
\end{figure}

In \reffi{fig:UIFcomp2} we compare our result in each of the three
schemes discussed above, i.e.\ the ``$\Ab,\;\tsb$~OS'', the 
``$\mb$~\drbar'' and the ``$\Ab$, $\tsb$ \drbar'' schemes, with the result of
\citere{mhiggsEP4}. The difference $\De\Mh$ between our result and the 
result of \citere{mhiggsEP4} is shown for each of the three schemes as a 
function of $\tb$  for $\mgl = 1500 \gev$, $\mu = -1000 \gev$, and 
$\MA = 700 \gev$. 
The differences stay below $1 \gev$ for $\tb \lsim 50$, where our result
in the ``$\Ab$, $\tsb$ \drbar'' scheme shows the biggest deviation from the
result of \citere{mhiggsEP4}, while as expected, the difference is
smallest for the ``$\Ab$, $\tsb$ OS'' scheme. For $\tb > 50$ large
deviations occur because of the sharp decrease of $\Mh$ in this region
(see e.g.\ \reffi{fig:mhtb_MAlarge_MUneg}).


\section{Conclusions}
\label{sec:conclusions}

We have obtained results for the two-loop \order{\alb\als} corrections 
to the neutral $\cp$-even Higgs-boson masses in the MSSM within
different renormalisation schemes. The leading $\tb$-enhanced
contributions of the $b/\Sbot$~sector can be incorporated into an 
appropriately chosen bottom
Yukawa coupling, for which we use the bottom-quark mass in the \drbar\
scheme with a resummation of the leading contributions. We have
analysed in detail the impact of the genuine (subleading)
\order{\alb\als}
two-loop corrections in different parts of the MSSM parameter space and
we have compared the results obtained in the different schemes.

We have shown that an on-shell scheme that is
frequently used in the $t/\Stop$~sector leads to numerically unstable
results if it is applied in the $b/\Sbot$~sector. The origin of the huge
corrections in this scheme was traced to the fact that it involves a
renormalisation condition 
for the sbottom mixing angle, $\tsb$, rather than for
the trilinear coupling, $\Ab$.

The other three schemes that we have analysed yield numerically
well-behaved and physically meaningful results. For $\mu > 0$ the effect
of the genuine \order{\alb\als} two-loop corrections is rather small,
typically of \order{100 \mev}. Corrections at this level
will nevertheless be relevant in view of the prospective accuracy of
measurements in the Higgs sector and of the top-quark mass at the ILC.
For $\mu < 0$ the effective bottom Yukawa coupling increases, leading to an
enhancement of the effects from the $b/\Sbot$~sector. While the
constraints arising from the measurement of the anomalous magnetic
moment of the muon favour the positive sign of $\mu$, it seems premature
at the present stage to discard the parameter region with $\mu < 0$.
For large values
of $\tb$ and $\mgl$ and large negative values of $\mu$ we find that the 
genuine \order{\alb\als} corrections can amount up to $3 \gev$. 
We have compared our result for the \order{\alb\als} corrections 
with the existing result in the literature, which was
obtained in the limit of $\tb \to \infty$, and found good agreement. 

The comparison of the results in the different schemes that we have
analysed and the investigation of the renormalisation scale dependence 
give an indication of the possible size of missing higher-order
corrections in the $b/\Sbot$~sector. For $\mu > 0$ the higher-order
corrections from the $b/\Sbot$~sector (beyond \order{\alb\als}) appear
to be sufficiently well 
under control even in view of the prospective ILC accuracy. 
This applies especially to the ``$\mb$ \drbar'' scheme, where 
the corrections beyond the improved one-loop result have been found to
be particularly small. For $\mu < 0$, on the other hand, sizable
higher-order corrections from the $b/\Sbot$~sector are possible. The
size of the individual corrections and also the difference between the 
analysed schemes varies significantly with the relevant parameters,
$\mu$, $\tb$, $\mgl$ and $\MA$. We estimate the uncertainty from missing
higher-order corrections in the $b/\Sbot$~sector to be about $2 \gev$ in 
this region of parameter space.

The results obtained will be implemented into the Fortran code
\fh~\cite{feynhiggs,feynhiggs2}. 
The evaluation of the results within the three schemes will allow to
obtain
an estimate of the size of the missing higher-order corrections
as a function of the chosen input parameters.


\subsection*{Acknowledgements}
We thank A.~Hoang, U.~Nierste, P.~Slavich and D.~St\"ockinger 
for helpful discussions. We are grateful to P.~Slavich for providing
the Fortran code for the ``$\msb$, $\Ab$ OS'' renormalisation scheme.
S.H.\ and G.W.\ thank the Max Planck Institut f\"ur
Physik, M\"unchen, for kind hospitality during part of
this work. G.W.\ thanks the CERN theory group for kind hospitality
during the final stage of this paper.
This work has been supported by the European Community's Human
Potential Programme under contract HPRN-CT-2000-00149 ``Physics at
Colliders''. 


\section*{Appendix: Counterterms of the quark/squark sector}
In section \ref{sec:ren} the counterterms have been given using the
definitions \eqref{transformation} and
\eqref{Sfermionmassenmatrixeigenwerte} for the sfermion masses and mixing
angles. In this appendix the counterterms are given in a more
general way allowing to use also other definitions for the sfermion masses and
mixing angles. Introducing a counterterm for the mixing angle needs a certain
choice of definitions of the sfermion masses and mixing angles. Instead of
using an explicit mixing angle counterterm the counterterm $\de
Y_{\tilde{q}}$ is introduced as 
\begin{align}
\de Y_{\tilde{q}} = (\cU_{\sq} \de \cM_{\sq} \cU^{\dagger}_{\sq})_{12}=
(\cU_{\sq} \de \cM_{\sq} \cU^{\dagger}_{\sq})_{21} ~, 
\end{align}
where the counterterm mass matrix $\de \cM_{\sq}$ contains the
counterterms of the parameters appearing in
\eqref{Sfermionmassenmatrix}. With the definitions \eqref{transformation} and 
\eqref{Sfermionmassenmatrixeigenwerte} $\de Y_{\tilde{q}}$ is
related to the mixing angle counterterm as follows   
\begin{align}
\de Y_{\tilde{q}} &= (\msqe^2 - \msqz^2)\, \de \tsq \,. 
\end{align}

\noindent
$\bullet$ {\bf Top  quark/squark sector:}

The counterterms for the top-quark mass \eqref{dmt} and the stop masses
\eqref{dmst} are already in a general form. The counterterm for the
mixing angle \eqref{ZusammenhangdeltathetadeltaM} is replaced by
\begin{align}
\de Y_{\tilde{t}} = 
\frac{1}{2}\Bigl(\re \Si_{\Stop_{12}}(m_{{\Stop}_{1}}^2)+\re 
\Si_{\Stop_{12}}(m_{{\Stop}_{2}}^2)\Bigr) ,
\end{align}
and the counterterm of the A-parameter \eqref{deltaAt} is rewritten as
\begin{align}
\de \At &=  \frac{1}{\mt}
 \Bigl[U_{\Stop_{11}} U_{\Stop_{12}}
 \bigl(\de \mste^2 - \de \mstz^2\bigr) 
      +(U_{\Stop_{11}} U_{\Stop_{22}} + U_{\Stop_{12}} U_{\tilde{t}_{21}})\,
      \de Y_{\tilde{t}} 
   - \de \mt \, ( \At - \mu \cot \be)  \Bigr]~.
\end{align}

\noindent
$\bullet$ {\bf Analogous to the top quark/squark sector:}

As in the top quark/squark sector the counterterm for the mixing angle
\eqref{dthetab} is replaced by 
\begin{align}\label{dYbanalogtop} 
\de Y_{\tilde{b}} = 
\frac{1}{2} \Bigl(\re \Si_{\tilde{b}_{12}}(\msbe^2)
     +\re \Si_{\tilde{b}_{12}}(\msbz^2) \Bigr)\; .
\end{align} 
The dependent counterterms of the  $\tilde{b}_1$-mass \eqref{ms1CT}
and of the A-parameter \eqref{Abparameter} are rewritten as follows:
\begin{align} \non
\de \msbe^2 &= \frac{1}{U_{\Sbot_{11}}^2}\bigl[
- U_{\Sbot_{12}}^2 \de\msbz^2 
+ 2 U_{\Sbot_{12}} U_{\Sbot_{22}}
    \de Y_{\tilde{b}}
   + U_{\Stop_{11}}^2 \de \mste^2   
   + U_{\Stop_{12}}^2 \de \mstz^2  
\\[1.5mm] & \quad\ 
  - 2 U_{\Stop_{12}} U_{\Stop_{22}} \de Y_{\tilde{t}} 
   + ( 2 \mb \de \mb - 2 \mt \de \mt)\bigr]~,\label{dmsb1allg} \\[1.5mm]\non
\de \Ab &= \frac{1}{\mb}\Bigl[
-\frac{U_{\Sbot_{12}}}{U_{\Sbot_{11}}} \de \msbz^2 +
 \frac{U_{\Sbot_{22}}}{U_{\Sbot_{11}}} \de Y_{\tilde{b}}
- \de m_{b} ( A_{b} - \mu \tb 
             - 2 \frac{U_{\Sbot_{12}}}{U_{\Sbot_{11}}} \mb) \\[1.5mm]
&  \quad\ \quad\
+ \frac{U_{\Sbot_{12}}}{U_{\Sbot_{11}}} \bigl( U_{\Stop_{11}}^2 \de \mste^2
 + U_{\Stop_{12}}^2 \de \mstz^2 
 - 2 U_{\Stop_{12}} U_{\Stop_{22}} \de Y_{\tilde{t}}
-2 \mt \de \mt \bigr) \Bigr]~. 
\end{align}

\noindent
$\bullet$ {\bf \boldmath{\drbar} bottom-quark mass}

The A-parameter counterterm \eqref{AbcountertermDR} is written in the
following way
 \begin{align} \non
 \de \Ab &= 
 \frac{1}{\mb}\Bigl[
 -\frac{U_{\Sbot_{12}}}{U_{\Sbot_{11}}} 
   \re \Si_{\Sbot_{22}}^{\rm div}(\msbz^2)
 + \frac{U_{\Sbot_{22}}}{2 U_{\Sbot_{11}}}
  \bigl(\re \Si_{\Sbot_{12}}^{\rm div}(\msbe^2)+
   \re \Si_{\Sbot_{12}}^{\rm div}(\msbz^2)\bigr)\\[1.5mm]
& \non \quad \quad\ 
  + \frac{U_{\Sbot_{12}}}{U_{\Sbot_{11}}}\Bigl( U_{\Stop_{11}}^2  
  \re  \Si_{\Stop_{11}}^{\rm div}(\mste^2)
  + U_{\Stop_{12}}^2 
  \re  \Si_{\Stop_{22}}^{\rm div}(\mstz^2) \\[1.5mm]
& \non \quad \quad\ 
 -  U_{\Stop_{12}} U_{\Stop_{22}} 
  \bigl(\re \Si_{\Stop_{12}}^{\rm div}(\mste^2)+
 \re \Si_{\Stop_{12}}^{\rm div}(\mstz^2)\bigr)
\\[1.5mm] & \non \quad \quad\
  -  \mt^2 (\re {\Si}_{{t}_L}^{\rm div}(\mt^2) + 
  \re {\Si}_{{t}_R}^{\rm div}(\mt^2) + 
 2\re {\Si}_{{t}_S}^{\rm div}(\mt^2))
\Bigr) \Bigr]
  + \edz  (2
  \frac{U_{\Sbot_{12}}}{U_{\Sbot_{11}}} \mb 
\\[1.5mm] &  \quad \quad
- A_{b} + \mu\tb)  \Bigl 
  (\re {\Si}_{{b}_L}^{\rm div} (\mb^2)
 + \re {\Si}_{{b}_R}^{\rm div} (\mb^2) 
 + 2\re {\Si}_{{b}_S}^{\rm div} (\mb^2) \Bigr)~,
 \end{align} 
avoiding an explicit definition of the mixing angles.
The dependent counterterm for the mixing angle \eqref{thetabinAbMBdrbar}
is replaced by
\begin{align}\non
 \de Y_{\tilde{b}} &= 
  \frac{U_{\Sbot_{11}}}{U_{\Sbot_{22}}}\mb \de \Ab 
 +\frac{U_{\Sbot_{11}}}{U_{\Sbot_{22}}} \de\mb
  ( \Ab - \mu \tb - 2 \frac{U_{\Sbot_{12}}}{U_{\Sbot_{11}}} \mb)
+\frac{U_{\Sbot_{12}}}{U_{\Sbot_{22}}} \bigl[\de \msbz^2\\[1.5mm]
& \quad\ 
 - U_{\Stop_{11}}^2 \de\mste^2
 - U_{\Stop_{12}}^2 \de\mstz^2 
 + 2 U_{\Stop_{12}} U_{\Stop_{22}}  \de Y_{\tilde{t}} 
 + 2 \mt \de \mt \bigr]~, 
\end{align}
and the counterterm for the $\tilde{b}_1$-mass \eqref{msb1inAbMBdrbar} by
\begin{align}\non
\de \msbe^2 &= 
 \frac{1}{U_{\Sbot_{11}}^2} \Bigl[ 
(1 - 2 U_{\Sbot_{12}}^2)
 \Bigl(U_{\Stop_{11}}^2 \de \mste^2 
      +U_{\Stop_{12}}^2 \de \mstz^2 
      -2 U_{\Stop_{12}} U_{\Stop_{22}} \de Y_{\tilde{t}} 
 - 2 \mt \de \mt \Bigr)
 + U_{\Sbot_{12}}^2 \de \msbz^2
\\[1mm] &  \quad 
+ 2 U_{\Sbot_{11}} U_{\Sbot_{12}} \mb \de \Ab
+ \Bigl( 2 U_{\Sbot_{12}} U_{\Sbot_{11}}
       ( \Ab - \mu \tb) + 2 (1 - 2
 U_{\Sbot_{12}}^2) \mb \Bigr) \de \mb \Bigr]~. 
\end{align}

\noindent
$\bullet$ {\bf \boldmath{\drbar} mixing angle and \boldmath{$\Ab$}}

The counterterm for the mixing angle \eqref{thetadrbar} is replaced by
\begin{align}
 \de Y_{\tilde{b}} = 
 \frac{1}{2}\bigl(\re \Si_{\Sbot_{12}}^{\rm div}(\msbe^2)+
       \re \Si_{\Sbot_{12}}^{\rm div}(\msbz^2)\bigr) \; .
\end{align}
The dependent counterterm for the bottom quark mass \eqref{deltambabh}
is rewritten as the following combination of counterterms:
\begin{align}\non
\de \mb &= - \Bigl[\mb\, \de \Ab - 
 \frac{U_{\Sbot_{22}}}{U_{\Sbot_{11}}}  \de Y_{\tilde{b}}
 + \frac{U_{\Sbot_{12}}}{U_{\Sbot_{11}}} \de \msbz^2 
- \frac{U_{\Sbot_{12}}}{U_{\Sbot_{11}}} \Bigl( 
 U_{\Stop_{11}}^2 \de \mste^2   
 + U_{\Stop_{12}}^2 \de \mstz^2  
 - 2 U_{\Stop_{12}} U_{\Stop_{22}} \de Y_{\tilde{t}}   
\\[1.5mm] & \quad\ 
 - 2 \mt\, \de \mt\Bigr) \Bigr]
 \Bigl[\Ab -\mu \tb - 2 \mb \frac{U_{\Sbot_{12}}}{U_{\Sbot_{11}}} \Bigr]^{-1}
 ~. \label{mbinAbthetabdrbarallg}
\end{align}
The counterterm for the $\tilde{b}_1$-mass is obtained by inserting the
expression  \eqref{mbinAbthetabdrbarallg} for the bottom quark mass into
the expression \eqref{dmsb1allg}.\\[1cm]

\noindent
$\bullet$ {\bf On-shell mixing angle and \boldmath{$\Ab$} }

The renormalised vertex \eqref{lambdahut} has the following general form
\begin{align}\non
 \hat{\La}(p_A^2, p_{\Sbote}^2, p_{\Sbotz}^2) &= \La (p_A^2,
 p_{\Sbote}^2, p_{\Sbotz}^2) + \frac{i e}{2 M_W \sin
 \theta_W}(U_{\Sbot_{11}}U_{\Sbot_{22}}- U_{\Sbot_{12}}U_{\Sbot_{21}}) \Bigl[
 \mb \tb \de \Ab  \\ & \quad  + (\mu + \tb \Ab)\Bigl( \de \mb +
 \frac{1}{2} \mb (\de
 Z_{\Sbote  \Sbote} + \de Z_{\Sbotz \Sbotz})\Bigr)\Bigr]~.
\end{align}
Using the renormalisation condition \eqref{lambdahutbed} the
counterterms of the A-parameter and the bottom quark mass can be derived
as
\begin{align} 
\non
\de \Ab &= 
  i \frac{\MW \sin \theta_W}{e\, \mb \tb (U_{\Sbot_{11}} U_{\Sbot_{22}}
    - U_{\Sbot_{12}} U_{\Sbot_{21}})} 
 \Bigl( \La(0, \msbe^2, \msbe^2) + \La(0, \msbz^2, \msbz^2) \Bigr)
\\[1.5mm]& \quad \non 
- \frac{\mu + \Ab \tb}{2 \tb}
  (\de Z_{\Sbote  \Sbote}  + \de Z_{\Sbotz \Sbotz}) 
\\ & \quad \non 
  - \frac{\mu + \Ab \tb}{\mb \tb}\Bigg[ 
 i \frac{\MW \sin \theta_W}{e \tb (U_{\Sbot_{11}} U_{\Sbot_{22}}-
   U_{\Sbot_{12}} U_{\Sbot_{21}})} 
  \Bigl(\La(0, \msbe^2,\msbe^2) + \La(0, \msbz^2, \msbz^2)\Bigr) \\[1.5mm]
& \qquad \non
  - \frac{\mb(\mu + \Ab  \tb)}{2 \tb} 
   (\de Z_{\Sbote  \Sbote}  + \de Z_{\Sbotz \Sbotz})  
+\frac{U_{\Sbot_{12}}}{U_{\Sbot_{11}}} \de  \msbz^2
 - \frac{U_{\Sbot_{22}}}{U_{\Sbot_{11}}} \de Y_{\tilde{b}} 
  -\frac{U_{\Sbot_{12}}}{U_{\Sbot_{11}}} \Bigl(
   U_{\Stop_{11}}^2 \de \mste^2
\\[1.5mm]& \qquad 
  + U_{\Stop_{12}}^2 \de \mstz^2    
  -2 U_{\Stop_{12}} U_{\Stop_{22}} \de Y_{\tilde{t}}
 - 2 \mt \de \mt \Bigr)
 \Bigg] 
\KKL \mu\Bigl(\tb + 
 \frac{1}{\tb}\Bigr) 
 + 2 \mb \frac{U_{\Sbot_{12}}}{U_{\Sbot_{11}}}
  \KKR^{-1}~,
\end{align}
and
\begin{align} \non
\de \mb &=  \Bigl[
  i \frac{\MW \sin \theta_W}{e\,\tb (U_{\Sbot_{11}} U_{\Sbot_{22}}-
   U_{\Sbot_{12}} U_{\Sbot_{21}})}
  \Bigl(\La(0, \msbe^2, \msbe^2) + \La(0, \msbz^2, \msbz^2) \Bigr) 
 \\[1.5mm]& \quad\ \non
  - \frac{\mb(\mu + \Ab \tb)}{2 \tb} 
  (\de Z_{\Sbote  \Sbote}  + \de Z_{\Sbotz \Sbotz}) 
  -\frac{U_{\Sbot_{22}}}{U_{\Sbot_{11}}} \de Y_{\tilde{b}} +
  \frac{U_{\Sbot_{12}}}{U_{\Sbot_{11}}} \de \msbz^2 
-  \frac{U_{\Sbot_{12}}}{U_{\Sbot_{11}}}
  \Bigl( U_{\Stop_{11}}^2 \de \mste^2 
 \\[1.5mm] & \qquad\ 
 + U_{\Stop_{12}}^2 \de \mstz^2    
 - 2 U_{\Stop_{12}} U_{\Stop_{22}}\de Y_{\tilde{t}} 
  - 2  \mt \de \mt 
\Bigr) 
\Bigr]
\Bigl[
  \mu\Bigl(\tb + \frac{1}{\tb}\Bigr) 
  +2 \mb \frac{U_{\Sbot_{12}}}{U_{\Sbot_{11}}}
  \Bigr]^{-1} ~,
  \end{align}
replacing  \eqref{dAbUIFartig} and \eqref{dmbmitthetauAbonshell}.
The counterterm of the mixing angle \eqref{mixangleUIF} is replaced by 
\eqref{dYbanalogtop}.


\end{document}